\documentclass[prd,showpacs,aps,nofootinbib,floatfix,floats,superscriptaddress,twocolumn]{revtex4-1} 

\usepackage[T1]{fontenc}
\usepackage[utf8]{inputenc}
\usepackage[english]{babel}
\usepackage{url}
\usepackage{mathrsfs}
\usepackage{amsfonts}
\usepackage{amsmath}
\usepackage{amssymb}
\usepackage{amsthm}
\usepackage{graphicx}

\usepackage[toc,title]{appendix}
\usepackage{bm}
\usepackage{booktabs}
\usepackage{braket}
\usepackage{caption}
\usepackage{enumerate}
\usepackage{enumitem}
\usepackage[Bjornstrup]{fncychap} 
\usepackage{fnpos} %
\usepackage[bottom,multiple]{footmisc} 
\usepackage{lipsum}	
\usepackage{mathtools} 
\usepackage[]{natbib}	
\usepackage{quoting}
\quotingsetup{font=small}
\usepackage{sidecap}
\usepackage{subfig}
\usepackage{textcomp}
\usepackage{tikz}
\usetikzlibrary{decorations.pathmorphing,calc}
\usepackage{titlesec}
\usepackage{tensor}
\usepackage{xfrac}

\usepackage{csquotes}
\usepackage{helvet}
\usepackage{listings}

\usepackage[bookmarks,bookmarksopen=true,colorlinks,citecolor=blue,linkcolor=blue,urlcolor=blue]{hyperref}

\makeFNbelow
\newcommand{\be}{\begin{equation}}
\newcommand{\ee}{\end{equation}}
\newcommand{\bn}{\begin{equation*}}
\newcommand{\en}{\end{equation*}}

\DeclarePairedDelimiter{\abs}{\lvert}{\rvert}	

\theoremstyle{plain}

\newcommand{\AEI}{\affiliation{Max Planck Institute for Gravitational Physics (Albert Einstein Institute), Am M\"uhlenberg 1, Potsdam-Golm, 14476, Germany}}
\newcommand{\SISSA}{\affiliation{SISSA, International School for Advanced Studies, Via Bonomea 265, 34136 Trieste, Italy}}
\newcommand{\INFN}{\affiliation{INFN Sezione di Trieste, Via Valerio 2, 34127 Trieste, Italy}}

\begin{document}

\rmfamily
\title{Quasinormal modes of weakly charged Einstein-Maxwell-dilaton black holes}
\author{Costantino Pacilio}\thanks{cpacilio@sissa.it}  \SISSA \INFN
\author{Richard Brito}\thanks{richard.brito@aei.mpg.de}  \AEI 
\date{\today}

\begin{abstract}
Einstein-Maxwell-dilaton theory is an interesting and well-motivated theoretical laboratory to explore the impact of new fundamental degrees of freedom in the context of testing the no-hair conjecture, due to the existence of hairy black hole solutions together with the propagation of scalar, vector and tensor modes. In this paper we compute the quasinormal mode spectrum of static and slowly rotating black holes for generic values of the dilaton coupling, within a weak electric charge approximation. Our results suggest that these spacetimes are stable for generic values of the dilaton coupling and the black hole charge. We also show that while gravitational modes are only weakly affected by the coupling with the dilaton, the spectrum of electromagnetic modes exhibits a more pronounced dilaton-dependent breaking of isospectrality between the axial and polar sectors. We further show that the gravitational quasinormal modes are well approximated by the properties of unstable null circular geodesics in those spacetimes, while the treatment of electromagnetic and scalar modes can be simplified by a suitably modified Dudley-Finley scheme for the perturbed equations.
\end{abstract}
\maketitle
%
\section{Introduction}
With the advent of gravitational-wave (GW) astronomy~\cite{Abbott:2016blz,Abbott:2016nmj,Abbott:2017vtc,Abbott:2017gyy,Abbott:2017oio,TheLIGOScientific:2017qsa}, we are now in a unique position to test Einstein's theory of General Relativity (GR) to unprecedented levels~\cite{testing_2015,TheLIGOScientific:2016src,Yunes:2016jcc,Berti:2018cxi,berti_2018_2}. One of the most striking predictions of GR is that the unique stationary and asymptotically flat black hole (BH) solution in vacuum is described by the Kerr geometry~\cite{Israel:1967wq,Carter:1971zc,Hawking:1971vc,Robinson:1975bv}. This remarkable fact implies that, to a very good approximation, all BHs in the Universe are expected to be uniquely described by their mass and spin. This result is commonly referred as the ``no-hair hypothesis''~\cite{cardoso_gualtieri}. Similar no-hair theorems have been obtained also in the context of modified gravity, most notably in some classes of scalar-tensor theories~\cite{herdeiro_radu,sotiriou_2015}. On the other hand, BHs \emph{do} have hair in several extended theories, such as Einstein-dilaton-Gauss-Bonnet gravity, dynamical Chern-Simons gravity, massive (bi)gravity and Lorentz violating gravity~\cite{testing_2015, sotiriou_2015,Babichev:2015xha,barack_2018}.

Testing the validity of the no-hair hypothesis is one of the most exciting prospects of future GW detectors~\cite{cardoso_gualtieri,berti_2018_2}. When slightly perturbed, BHs relax down to equilibrium through the emission of GWs described by a set of exponentially damped sinusoids with specific frequencies and damping times, the so-called quasinormal modes (QNMs), which dominate the late stages of the GW signal emitted from a binary BH merger~\cite{Berti:2009kk,berti_2018_2}. For a Kerr BH in GR, the entire QNM spectrum is uniquely determined by the BH's spin and mass. Therefore the detection of multiple QNM frequencies from the remnant of a BH merger would allow us to perform null tests of the no-hair hypothesis and possibly to detect deviations from the Kerr geometry~\cite{Dreyer:2003bv,Berti:2005ys,Gossan:2011ha,Meidam:2014jpa,Yang:2017zxs,Brito:2018rfr,berti_2018_2}. 

These experimental prospects call for a theoretical effort to compute the QNMs of BHs in alternative theories of gravity. This can be done both in a parametrized theory-agnostic fashion \cite{post_kerr,tattersall_2018} or case by case for specific examples of modified theories  of gravity. Most notably, QNM frequencies for BHs in theories beyond GR were computed for spherically symmetric solutions in theories such as Einstein-Maxwell-dilaton~\cite{Konoplya:2001ji,ferrari_1} (for a small subset of the theory), in dynamical Chern-Simons gravity~\cite{Molina:2010fb}, Einstein-dilaton-Gauss-Bonnet gravity~\cite{Pani:2009wy,Blazquez-Salcedo:2016enn,Blazquez-Salcedo:2017txk} and in massive (bi)gravity~\cite{Brito:2013wya,Brito:2013yxa,Babichev:2015zub}. 
The extension to spinning BHs, beyond the known QNMs of Kerr BHs in GR, has been mostly hindered by the lack of known exact analytical solutions in modified theories of gravity and by the generic difficulty to separate the equations of motion describing perturbations in those theories when dealing with axisymmetric spacetimes.  The only remarkable exception to this rule is the Kerr-Newman case in Einstein-Maxwell theory, for which QNMs have only recently been computed~\cite{berti_slow_1,berti_slow_2,Mark:2014aja,Dias:2015wqa}. Due to these difficulties, most of the estimates for spinning BHs have made use of the connection between the light ring and QNMs~\cite{Blazquez-Salcedo:2016enn,post_kerr,lehner_1}. However, this connection is formally valid only in the eikonal limit and fails to fully describe all the properties of the QNMs when non-minimal couplings to additional degrees of freedom are present~\cite{ferrari_1,Blazquez-Salcedo:2016enn,Blazquez-Salcedo:2017txk,Molina:2010fb}. In fact, unlike static perturbations of Schwarzschild and Reissner-Nordstr\"{o}m BHs in GR, which have the remarkable property that the axial and polar sectors are isospectral~\cite{chandra_book}, isospectrality is easily violated in alternative theories of gravity~\cite{ferrari_1,Blazquez-Salcedo:2016enn,Blazquez-Salcedo:2017txk,Molina:2010fb}, a feature that is not predicted by the light ring approximation.

In addition, jointly with the efforts to compute inspiral waveforms~\cite{Mirshekari:2013vb,Lang:2013fna,Lang:2014osa,Bernard:2018hta,Julie:2017pkb,Julie:2017ucp,Julie:2017rpw} and with the recent first numerical relativity simulations of binary BH mergers in theories beyond GR~\cite{Healy:2011ef,Berti:2013gfa,Okounkova:2017yby,lehner_2}, an accurate knowledge of the QNMs in those theories is necessary for the long-term prospect of building accurate inspiral-merger-ringdown waveform models in alternative theories of gravity, a missing piece at our disposal that may be crucial to dig up very small deviations from GR from the data. 

Within this context, in this work we compute the QNMs for non-spinning and slowly-spinning BH solutions of Einstein-Maxwell-dilaton (EMD) theory, a theory that has recently received a lot of interest. This theory is a good proxy for studies of more generic theories of gravity, mainly due to the fact that it admits hairy BH solutions and the presence of scalar and vector modes, in addition to the tensor (metric) modes. The dynamics of binary BHs in EMD theories was recently studied both numerically~\cite{lehner_2} and in a post-Newtonian expansion~\cite{Julie:2017rpw}, while analytical estimates of the merger properties were given in~\cite{lehner_1}. On the other hand, aside from the limiting case of the Kerr-Newman and Reissner-Nordstr\"{o}m solutions, QNMs in this theory were only computed for static BHs in a particular subset of the theory~\cite{Konoplya:2001ji,ferrari_1}. The main goal of this work is to extend this computation to a much wider parameter space of EMD. 

The action describing EMD that we will consider is given by~\cite{solution_3} 
\be
\label{eq:action}
S=\int d^4x\frac{\sqrt{-g}}{16\pi}\left[R-2g^{ab}\nabla_a\Phi\nabla_b\Phi-e^{-2\eta\Phi}F_{ab}F^{ab}\right]\,,
\ee
where $\Phi$ is a real scalar field (the dilaton), while $F_{ab}=\partial_aA_b-\partial_bA_a$ is the Maxwell tensor of the real electromagnetic (EM) potential $A_a$ (which does not need to coincide with the photon field of the Standard Model of particle physics)\footnote{The action~\eqref{eq:action} does not necessarily describe a modified theory of gravity \emph{per se}, since the vector and dilaton fields are minimally coupled to gravity, however one can show that, through a conformal transformation of the metric, this theory also admits an equivalent Jordan frame where the dilaton couples non-minimally to gravity (see e.g.~\cite{lehner_2}).}. The action depends upon the parameter $\eta$, the dilaton coupling. When $\eta=0$, it reduces to the Einstein-Maxwell action and the BH solutions reduce to the Kerr-Newman family. The case $\eta=1$, corresponds to a low-energy limit of string theory, while $\eta=\sqrt{3}$ can be obtained via a four-dimensional compactification of the five-dimensional Kaluza-Klein theory \cite{solution_1,solution_3}. In the following we will not restrict ourselves to any of these values and instead consider $\eta$ to be a free parameter of the theory.

\subsection{Executive summary}
\label{sec:introduction}
For the reader’s convenience, we summarize here the structure of the paper and our main results.

In Sec.~\ref{sec:formalism} we introduce the static and slowly rotating BH solutions of EMD. We quantify the notion of ``weak electric charge'' that we will use through this work, in terms of the charge-to-mass ratio $v=Q/M$, with $Q$ the BH charge and $M$ its mass, showing that the notion breaks down above a given $\eta$: for each $v$, we are limited in our approximation to a maximum value of $\eta$ where we expect our approximation to break down [cf. Eq.~\eqref{eq:thumb}].
We derive the equations of motion (EOM) describing generic small perturbations of the static BH solutions. Besides gravitational and EM waves, already present in ordinary Einstein-Maxwell theory, the dilaton induces also the propagation of scalar waves. Details on the derivation are shown in Appendix \ref{appendix:1}, where we also elucidate some subtleties in the derivation, connected with the weak-charge approximation.

Sec.~\ref{sec:qnmgeral} is devoted to the computation of the QNMs. We first compute them analytically using the geodesic correspondence~\cite{pani_lyapunov} in Sec.~\ref{sec:eikonal}, which provides rough estimates of the leading gravitational modes in the weak charge limit: in particular, we show that the modes depend very weakly on $\eta$, in agreement with the numerical results of~\cite{lehner_2}. In Sec.~\ref{sec:qnm} we numerically compute the gravitational, EM and scalar QNMs for static BHs in EMD, using Leaver's continued fraction (CF) method. In doing so, we extend the computation of QNMs in EMD beyond the particular values $\eta=0$ and $\eta=1$, the only cases treated generically in the literature so far~\cite{leaver_charge,andersson,chandra_book,ferrari_1}, and the computation of Ref.~\cite{Konoplya:2001ji} that computed the QNMs for generic $\eta$ but only for the axial sector. We explore the multipoles $l=0,1,2,3$ without finding any unstable mode. We confirm that both polar and axial gravitational modes depend only weakly on $\eta$ and that they are nearly isospectral, the breaking of isospectrality (ISO-breaking) being at most of the order of the percent for $v\approx 0.6$. On the other hand, isospectrality can be significantly broken for the EM modes. In particular, we find that the ISO-breaking for EM modes can reach absolute values of the order of $10\%$.

To elucidate these results we employ the so-called Dudley-Finley (DF) approximation in Sec.~\ref{sec:df}, which is particularly appropriate when the backreaction of matter fields on the vacuum geometry is small. In its original version, the DF approximation consists in perturbing the dynamical fields independently from each other. We devise a modified DF scheme adapted to EMD: in our scheme, one perturbs the metric as an independent field, but the vector and the dilaton fields are perturbed together, keeping on their mutual coupling. We find a good qualitative and quantitative agreement with the fully coupled results of Sec.~\ref{sec:qnm}. Our DF scheme captures the essential physics behind ISO-breaking in the EM sector, also providing a simpler way for its estimation.

The DF approximation becomes particularly useful in Sec.~\ref{sec:rotation}, where we turn our attention to slowly rotating BHs. Indeed, the EOM for perturbed rotating BHs are notoriously difficult to deal with, both at the analytical and at the numerical level. We compute the corrections due to the BH spin to the axial gravitational QNMs using the full perturbed equations, showing that they are consistent with the Kerr-Newman QNMs obtained in Refs.~\cite{berti_slow_1,berti_slow_2,Mark:2014aja,Dias:2015wqa} in the limit $\eta=0$. Finally, we resort to the DF approximation to compute the corrections to the EM modes both in the axial and in the polar sector, finding that our previous conclusions on ISO-breaking are left substantially unchanged.

We use the mostly plus $(-+++)$ metric convention and use $G=c=1$ units.  To facilitate comparison with our results, we made many of our calculations available online as {\scshape Mathematica}\textsuperscript{\textregistered} notebooks~\cite{webpages}.
\section{Framework}
\label{sec:formalism}
Our starting point is the EMD action given in Eq.~\eqref{eq:action}. Varying this action with respect to the different degrees of freedom leads to the EOM given by
\begin{subequations}
\begin{align}
&S\equiv\nabla^a\nabla_a\Phi+\frac{\eta}{2}e^{-2\eta\Phi}F_{ab}F^{ab}=0\,,\\
&J_a\equiv\nabla_b\left(e^{-2\eta\Phi}F\indices{^b_a}\right)=0\,,\\
&E_{ab}\equiv G_{ab}-T^\Phi_{ab}-T^F_{ab}=0\,,
\end{align}
\end{subequations}
where $G_{ab}=R_{ab}-g_{ab}R/2$ is the Einstein tensor. The scalar and EM stress-energy tensors are respectively
\begin{subequations}
\begin{align}
&T^\Phi_{ab}=2\nabla_a\Phi\nabla_b\Phi-(\nabla\Phi)^2g_{ab}\,,\\
&T^F_{ab}=e^{-2\eta\Phi}\left(2F_{ac}F\indices{_b^c}-\frac{1}{2}F^2g_{ab}\right)\,,
\end{align}
\end{subequations}
where we used the shorthand notations $(\nabla\Phi)^2=g^{ab}\nabla_a\Phi\nabla_b\Phi$ and $F^2=F_{ab}F^{ab}$.

\subsection{Black holes in EMD theory}
\label{sec:solutions}
\paragraph{Static black holes.} Static spherically symmetric BH solutions were derived in \cite{solution_1, solution_2}; see also \cite{solution_3}. They are electrically charged, and the scalar field presents a secondary hair.\footnote{By secondary hair we mean that the scalar hair is not a new independent charge, but it is a function of the mass $M$ and the electric charge $Q$ \cite{herdeiro_radu,Pacilio:2018gom}.} The line element is specified by
\begin{subequations}
\label{eq:line:1}
\begin{align}
&ds^2=-F(r)dt^2+\frac{dr^2}{F^(r)}+r^2G(r)\left(d\theta^2+\sin^2\theta d\phi^2\right)\,,\\
&F(r)=\left(1-\frac{R_+}{r}\right)\left(1-\frac{R_-}{r}\right)^{(1-\eta^2)/(1+\eta^2)}\,,\\
&G(r)=\left(1-\frac{R_-}{r}\right)^{2\eta^2/(1+\eta^2)}\,.
\end{align}
\end{subequations}
where $R_+$ and $R_-$ are, respectively, the radii of the outer and inner event horizons:
\begin{subequations}
\label{eq:rpm:1}
\begin{align}
&R_+=M\left(1+\sqrt{1-(1-\eta^2)\,v^2}\right)\,,\\
&R_-=M\left(\frac{1+\eta^2}{1-\eta^2}\right)\left(1-\sqrt{1-(1-\eta^2)\,v^2}\right)\,.
\end{align}
\end{subequations}
Here $M$ is the asymptotic mass and $v=Q/M$ is the electric charge-to-mass ratio. Finally, the scalar field and the vector potential are given by
\be
\label{eq:matter:1}
\Phi=\frac{\eta}{1+\eta^2}\log\left(1-\frac{R_-}{r}\right)\,,\,\,\,\,
A_a\,dx^a=\frac{Mv}{r}\,dt\,.
\ee
Notice that the EMD action is invariant under the reparametrization
\be
\label{eq:rep}
A_a\to e^{-\eta\Phi_0}A_a\,,\qquad \Phi\to\Phi -\Phi_0\,,
\ee
where $\Phi_0$ is a constant. This implies that the scalar field is specified up to a constant $\Phi_0$ and the electric charge up to a factor $e^{-\eta\Phi_0}$. For simplicity, we choose $\Phi_0$ so that $\Phi=0$ at spatial infinity.

When $\eta\neq 0$, $R_-$ is a true singularity; moreover spherical sections of coordinate radius $R_-$ have vanishing area, signaling a breakdown of the spacetime. In order to prevent naked singularities we must impose the reality of $R_+$ and $R_-$ and the inequality $R_-<R_+$, which result in an upper bound on the value of the electric charge (in this paper we exclude extremal BHs from our consideration):
\be
\label{eq:v:bound}
v^2<1+\eta^2.
\ee
The bound applies also to the case $\eta=0$, where $R_-$ becomes the radius of the inner Cauchy horizon.

The vector potential $A_a$ does not necessarily correspond to the photon field of the Standard Model and so standard arguments for the smallness of the electric charge do not necessarily apply~\cite{Cardoso:2016olt}\footnote{On the other hand, if the vector potential is identified with the photon field, one expects astrophysical BHs to have a small or vanishing electric charge \cite{charge_1,charge_2,charge_3}.}. However the approximation of small charge significantly simplifies the perturbed equations, and in some cases it was the only way we were able to compute the QNMs of these solutions using standard numerical methods. Therefore we will mostly work under this approximation. 

Since the first corrections to the metric occur at second order in the electric charge, we define the weak-charge limit as the expansion to order $\mathcal{O}(v^2)$. The approximate solution at $\mathcal{O}(v^2)$ reads
\begin{subequations}
\label{eq:line:2}
\begin{align}
&ds^2=-f(r)dt^2+\frac{dr^2}{f(r)}+r^2g(r)d\Omega^2\,,\\
&f(r)=1-\frac{2M}{r}+\frac{(1-\eta^2)\,M^2v^2}{r^2}\,,\\
&g(r)=1-\frac{\eta^2Mv^2}{r}
\end{align}
\end{subequations}
for the line element, and
\be
\label{eq:matter:2}
\Phi=-\frac{\eta\,M\,v^2}{2r}\,,\,\,\,\, A_a\,dx^a=\frac{Mv}{r}\,dt\,,
\ee
for the matter fields. The approximate outer and inner horizons are
\be
\label{eq:rpm:2}
R_+\simeq2M\left(1-\frac{(1-\eta^2)v^2}{4}\right)\,,\quad R_-\simeq\frac{M(1+\eta^2)v^2}{2}\,.
\ee
An observation is in order: we see from Eqs.\,\eqref{eq:line:1} and \eqref{eq:matter:1} that what we are actually expanding are the $v$-terms in $R_\pm/r$. If we require that our expansion be valid in the domain of outer communication, then it can fail at most for $r\sim M$. From \eqref{eq:rpm:1}, this shows that the quantity  $\abs{(1-\eta^2)v^2}$ must be small enough in order for the expansion to be consistent, as it is also clear from \eqref{eq:line:2} and \eqref{eq:rpm:2}. Therefore the small $v$ expansion effectively induces an upper limit on $\eta$. In practice, we will use the following rule of thumb: for each $v$, we consider the expansion meaningful if the inequality $\abs{(1-\eta^2)v^2}\leq0.5$ holds, i.e. if
\be
\label{eq:thumb}
0\leq\eta\leq\sqrt{1+0.5/v^2}\,.
\ee
For example, when $v=0.5$, \eqref{eq:thumb} gives $\eta\leq\sqrt{3}$.\\
\paragraph{Slowly rotating black holes.}
Fully rotating BH solutions in EMD theory have only been found for the particular case $\eta=\sqrt{3}$~\cite{frolov_metric,solution_3,lehner_1}. On the other hand, Ref.~\cite{solution_3} derived a generic expression for slowly rotating BHs with arbitrary values of $\eta$. The form of the metric is
\be
\label{eq:metric:slow}
ds^2=ds^2_\text{static}-2a\,\Omega(r)\sin^2\theta\, dtd\phi+\mathcal{O}(a^2)\,,
\ee
where
\be
\Omega(r)=\frac{2M}{r}+\frac{\left[r\eta^2-3M(1-\eta^2)\right]Mv^2}{3r^2}+\mathcal{O}(v^3)\,.
\ee
The slowly rotating vector potential is given by
\be
A_a\,dx^a=\frac{Mv}{r}\left(dt-a\sin^2\theta\,d\phi\right)+\mathcal{O}(a^2)\,,
\ee
while the dilaton field remains unchanged at linear order in $a$. The constant $a$ represents an ``unphysical spin parameter'' in the sense that it is not related to the angular momentum $J$ by the Kerr-Newman relation $J=aM$, but rather as~\cite[Eq.(36)]{solution_3}
\be
\label{eq:j}
J=a\,M\left(1+\frac{\eta^2\,v^2}{6}\right)+\mathcal{O}(v^3)\,.
\ee
We therefore define a ``physical spin parameter'' $a_J$ by
\be
\label{eq:aphys}
a_J=a\left(1+\frac{\eta^2\,v^2}{6}\right)\,,
\ee
such that $J=a_JM +\mathcal{O}(v^3)$. Finally, since in geometrized units the parameter $a_J$ has the dimension of a length, it is useful to re-scale it as $a_J=M\,\tilde{a}$ , in such a way as to isolate the universal mass scale $M$ and deal with a pure number $\tilde{a}$ in our approximation scheme.
\subsection{Perturbed equations of motion}
\label{sec:eq:pert}
We now consider generic linear perturbations of the BH solutions described above. For simplicity here we describe the derivation only for the static configuration. An independent derivation of the perturbed equations, but in a different gauge, was also partly presented in Ref.~\cite{Holzhey:1991bx}. The generalization to the spinning case is straightforward. All the perturbed equations, including the ones for the slowly rotating solution, are listed in a supplemental {\scshape Mathematica}\textsuperscript{\textregistered} notebook~\cite{webpages}.

We start by writing the perturbed metric as $g_{ab}=g_{ab}^{(0)}+\delta g_{ab}$, where $g_{ab}^{(0)}$ is given by the background line element \eqref{eq:line:1} and $\delta g_{ab}$ is the perturbation. Similarly $A_a=A_a^{(0)}+\delta A_a$ and $\Phi=\Phi^{(0)}+\delta\Phi$. In a spherically symmetric background the field perturbations can be decomposed in spherical harmonics with multipole number $l$ and azimuthal number $m$. This expansion naturally separates the perturbations into ``axial'' [which acquire a factor $\left(-1\right)^{l+1}$ under parity inversions] and ``polar'' [which instead acquire a factor $\left(-1\right)^l$]. The EOM for the perturbed fields are then solved in the Fourier domain, with complex frequency $\omega$. The monopole $l=0$ and the dipole $l=1$ require a separate, although simpler, treatment with respect to the higher multipoles. Here we concentrate on $l\geq2$, while for $l=0$ and $l=1$ we just spell below the final equations.

We expand the metric perturbations in tensor spherical harmonics in the standard Regge-Wheeler gauge \cite{regge_wheeler}
\be
\delta g_{ab}=\delta g^A_{ab}+\delta g^P_{ab}\,,
\ee
where hereafter superscripts $A$ and $P$ indicate axial and polar components respectively. They read
\begin{subequations}
\label{eq:metric:pert}
\begin{align}
&\delta g^A_{ab}=\sum_{l,m}\int d\omega\, e^{-i\omega t}\times \nonumber\\
&\times\begin{bmatrix}
0 &0 &-\frac{h_0(r)\partial_\phi Y_l^m}{\sin \theta} &h_0(r)\sin\theta\partial_\theta Y_l^m\\
\ast &0 &-\frac{h_1(r)\partial_\phi Y_l^m}{\sin \theta} &h_1(r)\sin\theta\partial_\theta Y_l^m\\
\ast &\ast &0 &0\\
\ast &\ast &\ast &0 
\end{bmatrix}\,,\\
&\delta g^P_{ab}=\sum_{l,m}\int d\omega\, e^{-i\omega t}Y_l^m\times \nonumber\\
&\times\begin{bmatrix}
f(r) H_0(r) &H_1(r) &0 &0\\
\ast &\frac{H_2(r)}{f(r)} &0 &0\\
\ast &\ast &r^2K(r) &0\\
\ast &\ast &\ast &r^2\sin^2\theta K(r) 
\end{bmatrix}\,,
\end{align}
\end{subequations}
where $Y_l^m\equiv Y_l^m(\theta,\phi)$ are the spherical harmonic functions and asterisks denote symmetrization. Similarly, we write the vector perturbations as
\be
\delta A_a=\delta A^A_a+\delta A^P_a\,,
\ee
with
\begin{subequations}
\label{eq:vector:pert}
\begin{align}
&\delta A^A_a=\sum_{l,m}\int d\omega\, e^{-i\omega t} \times \nonumber\\
&\times \left(0,0,-\frac{u_4(r)\partial_\phi Y_l^m}{\sin\theta},u_4(r)\sin\theta\partial_\theta Y_l^m\right)\,,\\
&\delta A^P_a=\sum_{l,m}\int d\omega\, e^{-i\omega t}\left(\frac{u_1(r)Y_l^m}{r},\frac{u_2(r)Y_l^m}{rf(r)},0,0\right)\,,
\end{align}
\end{subequations}
where we used the $U(1)$ gauge freedom of the vector potential to gauge out the angular components of $\delta A^P_a$. Since the vector potential appears in the EOM only through $F_{ab}$, it is more convenient to work with the components of the perturbed Maxwell tensor:
\be
\delta F_{ab}=\delta F^A_{ab}+\delta F^P_{ab}\,,
\ee
with
\begin{subequations}
\label{eq:f:pert}
\begin{align}
&\delta F^A_{ab}=\sum_{l,m}\int d\omega e^{-i\omega t}\times\nonumber\\
&\times\begin{bmatrix}
0 &0 &-\frac{i\omega u_4(r)\partial_\phi Y_l^m}{\sin\theta} & i\omega u_4(r)\sin\theta\partial_\theta Y_l^m\\
\ast &0 &\frac{u_4'(r)\partial_\phi Y_l^m}{\sin\theta} & -u_4'(r)\sin\theta\partial_\theta Y_l^m\\
\ast &\ast &0 &l(l+1)u_4(r)\sin\theta Y_l^m\\
\ast &\ast &\ast &0
\end{bmatrix}\,,\\
&\delta F^P_{ab}=\sum_{l,m}\int d\omega e^{-i\omega t}\times\nonumber\\
&\times\begin{bmatrix}
0 &f_{01}(r)Y_l^m &f_{02}(r)\partial_\theta Y_l^m &f_{02}(r)\partial_\phi Y_l^m\\
\ast &0 &f_{12}(r)\partial_\theta Y_l^m &f_{12}(r)\partial_\phi Y_l^m\\
\ast &\ast &0 &0\\
\ast &\ast &\ast &0
\end{bmatrix}\,,
\end{align}
\end{subequations} 
and the asterisks now denote anti-symmetrization. The radial functions $f_{01}$, $f_{02}$ and $f_{12}$ are expressed in terms of $u_1$ and $u_2$ as
\begin{subequations}
\label{eq:f:u}
\begin{align}
& f_{01}(r)=\frac{i r\,\omega u_2(r)+f(r)\left(r u_1'(r)-u_1(r)\right)}{r^2f(r)}\,,\\
& f_{02}(r)=\frac{u_1(r)}{r}\,,\\
& f_{12}(r)=\frac{u_2(r)}{r f(r)}\,.
\end{align}
\end{subequations}
They are not independent from each other, but they are related by the Bianchi identity
\be
\label{eq:bianchi}
f_{01}(r)=i\omega f_{12}(r)+f_{02}'(r)\,.
\ee
Finally, we write the perturbation of the scalar field as
\be
\label{eq:p:scalar}
\delta\Phi=\sum_{l,m}\int d\omega e^{-i\omega t}z(r)Y_l^m\,.
\ee

The derivation of the perturbed EOM is described in Appendix~\ref{appendix:1}. They are separable into axial EOM and polar EOM.\\
\paragraph{Axial EOM.} The axial group can be reduced to a system of two coupled second order differential equations,
\begin{subequations}
\label{eq:axial:eom}
\begin{align}
\begin{split}
&\text{``Maxwell equation'':}\\
&\left[\frac{d^2}{dr\indices{_\star^2}}+\omega^2\right]\hat{U}(r)=V_{FF}(r)\hat{U}(r)+V_{FK}(r)\hat{Q}(r)\,,\label{eq:p:1}
\end{split}\\
\begin{split}
&\text{``Regge-Wheeler equation'':}\\
&\left[\frac{d^2}{dr\indices{_\star^2}}+\omega^2\right]\hat{Q}(r)=V_{KK}(r)\hat{Q}(r)+V_{KF}(r)\hat{U}(r)\,,\label{eq:p:3}
\end{split}
\end{align}
\end{subequations}
where the functions $\hat{U}$ and $\hat{Q}$ are defined in Eq.\eqref{eq:redef:axial:3} of Appendix \ref{appendix:1}, and the potentials $V_{IJ}$'s are presented in a {\scshape Mathematica}\textsuperscript{\textregistered} notebook~\cite{webpages}. Here $r_\star$ is the tortoise coordinate $dr_\star/dr=1/F(r)$, which is defined in the domain $r_\star\in\left]-\infty,+\infty\right[$ for $r\in\left]r_+,+\infty\right[$.

We here choose to name the equations according to their limit when $v\to 0$. In particular they reduce respectively to the Maxwell equation (i.e. the equation describing EM perturbations) and to the Regge-Wheeler equation in the limit where the background reduces to a Schwarzschild BH. Via a linear transformation of the fields, Eqs.~\eqref{eq:axial:eom} can be put in the diagonal form
\be
\label{eq:axial:eom:2}
\left[\frac{d^2}{dr\indices{_\star^2}}+\omega^2\right]Z_i^A(r)=V_i^A(r)\,Z_i^A(r)\,\quad i=1,2
\ee
where the axial potentials $V^A_{1,2}$ are given by
\begin{widetext}
\begin{multline}
\label{eq:v:axial:1}
V_1^A(r)=\frac{(2 M-r) (6 M-l(l+1)\,r)}{r^4}+\\+\frac{M v^2 \left[M r \left(-27 \eta ^2+\left(7-9 \eta ^2\right)l(l+1)+4\right)+r^2 \left(\left(3 \eta ^2-2\right)l(l+1)+4\right)+6 \left(12 \eta ^2-7\right) M^2\right]}{3 r^5} + \mathcal{O}(v^4) \,,
\end{multline}
and
\begin{multline}
\label{eq:v:axial:2}
V_2^A(r)=\frac{l(l+1)(r-2 M)}{r^3}+\\+\frac{Mv^2\left[-M r \left(15 \eta ^2+9 \eta ^2\,l(l+1)+l(l+1)-20\right)+r^2 \left(3 \eta ^2+\left(3 \eta ^2+2\right)l(l+1)-4\right)+6 \left(3 \eta ^2-4\right) M^2\right]}{3 r^5}+\mathcal{O}(v^4)\,.
\end{multline}
\end{widetext}
\paragraph{Polar EOM.}
The polar perturbed EOM reduce to three coupled second order differential equations
\begin{widetext}
\begin{subequations}
\label{eq:polar:eom}
\begin{align}
\begin{split}
&\text{``Maxwell equation'':}\\
&\left[\frac{d^2}{dr\indices{_\star^2}}+\omega^2\right]\hat{F}(r)=V_{FF}(r)\hat{F}(r)+V_{FS}(r)\hat{S}(r)+V_{FK}(r)\hat{K}(r)+U_{FK}(r)\frac{d\hat{K}(r)}{dr_T}\,, \label{eq:p:1}
\end{split}\\
\begin{split}
&\text{``Scalar equation'':}\\
&\left[\frac{d^2}{dr\indices{_\star^2}}+\omega^2\right]\hat{S}(r)=V_{SS}(r)\hat{S}(r)+V_{SF}(r)\hat{F}(r)+U_{SF}(r)\frac{d\hat{F}(r)}{dr_T}\,, \label{eq:p:2}
\end{split}\\
\begin{split}
&\text{``Zerilli equation'':}\\
&\left[\frac{d^2}{dr\indices{_\star^2}}+\omega^2\right]\hat{K}(r)=V_{KK}(r)\hat{K}(r)+V_{KF}(r)\hat{F}(r)+V_{KS}(r)\hat{S}(r)\,, \label{eq:p:3}
\end{split}
\end{align}
\end{subequations}
\end{widetext}
where the functions $\hat{K}$, $\hat{F}$ and $\hat{S}$ are defined in Eqs.\eqref{eq:zerilli:redef} and \eqref{eq:zerilli:redef:2} of Appendix \ref{appendix:1}, and again the potentials $V_{IJ}$'s and $U_{IJ}$'s are presented in the {\scshape Mathematica}\textsuperscript{\textregistered} notebook~\cite{webpages}. The names of the equations are assigned in analogy with the axial case; in particular, Eq.\eqref{eq:p:2} reduces to the massless Klein-Gordon equation on a Schwarzschild background when $v\to0$. Contrary to the axial case, we were not able to find a simple way to diagonalize the system \eqref{eq:polar:eom}.\\

\paragraph{Lower multipoles.}
When $l=0$ the perturbed EOM reduce to the single scalar equation
\be
\label{eq:eom:0}
\left[\frac{d^2}{dr\indices{_\star^2}}+\omega^2\right]Z^0(r)=V^0(r)\,Z^0(r)\,,
\ee
where the potential $V_0$ is given by\footnote{For completeness we also computed the potential $V_0$ without employing the small-charge approximation. The result can be found in Appendix~\ref{appendix:3}.}
\begin{multline}
\label{eq:v0}
V^0(r)=\frac{2 M (r-2 M)}{r^4}\\-\frac{M^2 v^2 \left(6 \left[2 \eta ^2-1\right) M+\left(2-5 \eta ^2\right) r\right]}{r^5}+\mathcal{O}(v^4)\,.
\end{multline}
For $l=1$, the axial equation is
\be
\label{eq:eom:1a}
\left[\frac{d^2}{dr\indices{_\star^2}}+\omega^2\right]Z^{1,A}(r)=V^{1,A}(r)\,Z^{1,A}(r)\,,
\ee 
with
\begin{multline}
\label{eq:v1a}
V^{1,A}(r)=\frac{2(r-2 M)}{r^3}\\-\frac{M v^2 \left[\left(8-6 \eta ^2\right) M^2+\left(11 \eta ^2-6\right) M r-3 \eta ^2 r^2\right]}{r^5}+\mathcal{O}(v^4)\,.
\end{multline}
Finally, the $l=1$ polar equations are a system of two coupled differential equations
\begin{subequations}
\label{eq:eom:1p}
\begin{align}
&\left[\frac{d^2}{dr\indices{_\star^2}}+\omega^2\right]\hat{F}^1(r)=V_{FF}^1(r)\hat{F}^1(r)+V_{FS}^1(r)\hat{S}^1(r)\,,\\
&\left[\frac{d^2}{dr\indices{_\star^2}}+\omega^2\right]\hat{S}^1(r)=V_{SS}^1(r)\hat{S}(r)+V_{SF}^1(r)\hat{F}^1(r)\,.
\end{align}
\end{subequations}
The explicit expressions for the $l=1$ polar potentials are again provided in a {\scshape Mathematica}\textsuperscript{\textregistered} notebook~\cite{webpages}.

\section{QNMs of EMD Black Holes}
\label{sec:qnmgeral}
The perturbation equations derived above can be numerically solved with appropriate boundary conditions to compute the QNMs of the BH solutions presented in Sec.~\ref{sec:solutions}. However, before doing so, let us first estimate the expected QNMs using the geodesic correspondence~\cite{pani_lyapunov}. This approximation, commonly used in the literature, allows us to get analytical estimates for the QNM frequencies and is helpful to understand the behavior of the QNM frequencies when varying the parameters of the theory.
\subsection{The light ring approximation}
\label{sec:eikonal}
In the geodesic approximation,  which is formally only valid in the eikonal limit $l\gg1$, the frequencies and decay times of the QNMs can be derived from the properties of unstable circular null geodesics of the BH geometry. This connection was first explored in \cite{mashhoon_ferrari_1,mashhoon_ferrari_2,Mashhoon:1985cya} in the context of GR BHs and later formalized in \cite{pani_lyapunov} for generic BH spacetimes. For EMD BHs it was applied in Ref.~\cite{lehner_1} for the spinning BH solution when $\eta=\sqrt{3}$, and also for Kerr-Newman BHs in~\cite{Mashhoon:1985cya,Cardoso:2016olt}. 

We focus our attention to light rays orbiting in the equatorial plane $\theta=\pi/2$ which, in the eikonal correspondence, are related to the $l=\pm m\gg1$ QNMs (the plus and minus signs stand for corotating and counterrotating orbits respectively)~\cite{Yang:2012he}. Let then
\be
\dot{r}^2=V_\text{geo}(r)
\ee
be the equation describing the motion of a general null geodesic in the equatorial plane of the black hole (the dot denotes derivative with respect to the affine parameter). From \eqref{eq:line:1} and \eqref{eq:metric:slow} we get
\be
\label{eq:vgeo}
V_\text{geo}(r)=1-\frac{F(r)L^2}{r^2G(r)}\mp2a\frac{\Omega(r)L}{r^2G(r)}\,.
\ee
The equations of motion for the coordinates $t$ and $\phi$ are
\begin{subequations}
\begin{align}
&\dot{t}=\frac{1}{F(r)}\mp a\frac{\Omega(r)L}{r^2F(r)G(r)}\,,\\
&\dot{\phi}=\pm\frac{L}{r^2G(r)}+a\frac{\Omega(r)}{r^2F(r)G(r)}\,.
\end{align}
\end{subequations}
The Killing angular momentum $L$ and the radius $r_c$ of the innermost unstable circular orbit are determined by solving the system of the two equations $V_\text{geo}(r_c)=0$ and $V_\text{geo}'(r_c)=0$. Under small perturbations, unstable null geodesics (for which $V''_\text{geo}(r_c)>0$) decay with a principal rate $\Gamma_c$. Then, in the limit $l\gg1$, the BH QNMs are given by~\cite{pani_lyapunov}
\be
\label{eq:omega:eikonal}
\omega_n=\pm l\,\Omega_c-i\left(n+\frac{1}{2}\right)\Gamma_c\,,
\ee
where $n$ is the overtone number, $\Omega_c$ is the angular frequency of the orbit and $\Gamma_c$ is its dominant decay rate, given by
\be
\label{eq:gammac}
\Gamma_c=\sqrt{\frac{V_\text{geo}''(r_c)}{2\,\dot{t}^2}}\,.
\ee
Despite the fact that the geodesic correspondence is formally only valid in the eikonal limit $l\gg1$, it gives surprisingly accurate results even for low multipoles: in the Kerr-Newman case, one can verify that the real parts of the $l=2$ and $l=3$ fundamental gravitational modes are predicted within a few percents of accuracy (see e.g. Appendix A of~\cite{Cardoso:2016olt}). In particular, in the weakly charged limit we find
\begin{subequations}
\label{eq:omega:eikonal:2}
\begin{align}
&M\Omega_c=\frac{1}{3\sqrt{3}}\left(1+\frac{v^2}{6}\right)\pm\frac{2}{27}\,\tilde{a}\left(1+\frac{v^2}{2}\right)+\mathcal{O}(v^3,\tilde{a}^2)\label{eq:eikonal:3}\,,\\
&M\Gamma_c=\frac{1}{3\sqrt{3}}\left(1+\frac{v^2}{18}\right)\pm\frac{2\tilde{a}\,v^2}{243}+\mathcal{O}(v^3,\tilde{a}^2)\,,
\end{align}
\end{subequations}
where we have expressed the results in terms of the dimensionless spin parameter $\tilde{a}$. These expressions do not depend on $\eta$ (cf. Ref.~\cite{Mashhoon:1985cya} where the same results were obtained for Kerr-Newman). Therefore, using~\eqref{eq:omega:eikonal} and~\eqref{eq:omega:eikonal:2} to estimate the gravitational QNMs at small $l$, we end up with the prediction that they should depend very weakly on $\eta$ and that their value is very close to the Kerr-Newman one. 

This agrees with the estimates of Refs.~\cite{lehner_1} and the numerical simulations of Ref.~\cite{lehner_2}, which did not find any deviation from isospectrality within the numerical error. We shall see in Sec.~\ref{sec:qnm:gr}, by exactly computing the QNMs, that indeed the static gravitational QNMs depend very weakly on $\eta$ and that isospectrality is only mildly broken in the gravitational sector.

The estimates in Eq.~\eqref{eq:omega:eikonal:2} also show that the imaginary part of the QNM frequency depends very weakly on the spin, since we are expanding at both small $v$ and small $\tilde{a}$. Again, we shall see in Sec.~\ref{sec:rotation} that this holds when solving the perturbed EOM numerically. Moreover, we will show that~\eqref{eq:eikonal:3} provides a good quantitative estimate of the spin correction for the gravitational axial modes. 

\subsection{Computation of the static QNMs}
\label{sec:qnm}
We start by computing the QNMs in the static case. The QNM spectrum is obtained by solving the perturbed EOM with ingoing boundary conditions at the BH event horizon and outgoing boundary conditions at spatial infinity, forming an eigenvalue problem for the complex frequency $\omega$. Both axial and polar potentials vanish at the boundaries, implying that the generic wave function $Z(r)$ has the asymptotic behavior:
\be
Z(r)\sim
\begin{cases}
e^{-i\omega r_\star}\sim(r-R_+)^{-2\mu\,i\omega} &\text{for $r\to R_+$}\,,\\
e^{i\omega r_\star}\sim e^{-i\omega r}r^{2M\,i\omega} &\text{for $r\to+\infty$}\,.
\end{cases}
\ee
The constant $\mu$ can be obtained by integrating $dr_\star/dr=F^{-1}(r)$ close to the horizon, and it is equal to
\begin{multline}
\mu=M \left(\frac{\sqrt{1+\left(\eta ^2-1\right) v^2}-\eta ^2}{1-\eta ^2}\right)^{(\eta ^2-1)/(\eta ^2+1)}\times\\
\times\left(\frac{\sqrt{1+\left(\eta ^2-1\right) v^2}+1}{2}\right)^{2/(\eta ^2+1)}\,.
\end{multline}
Therefore a convenient ansatz for the wave functions is the following series expansion around the event horizon:
\begin{multline}
\label{eq:ansatz:1}
Z(r)=e^{-i\omega(r-R_+)}\,r^{2(M+\mu)\,i\omega}(r-R_+)^{-2\mu\,i\omega}\times\\
\times\sum_{k=0}^\infty a_k\,\left(1-\frac{R_+}{r}\right)^k\,.
\end{multline}
At order $\mathcal{O}(v^2)$, $\mu$ reduces simply to $M$, and the ansatz becomes
\begin{multline}
\label{eq:ansatz:2}
Z(r)=e^{-i\omega(r-R_+)}\,r^{4M\,i\omega}(r-R_+)^{-2M\,i\omega}\times\\
\sum_{k=0}^\infty a_k\,\left(1-\frac{R_+}{r}\right)^k\,.
\end{multline}
Since we work with coupled equations, we must define a vectorial wave function $\vec{Z}(r)$, with corresponding ansatz
\begin{multline}
\label{eq:ansatz:2}
\vec{Z}(r)=e^{-i\omega(r-r_+)}\,r^{4M\,i\omega}(r-R_+)^{-2M\,i\omega}\times\\
\sum_{k=0}^\infty\vec{a}_k\,\left(1-\frac{R_+}{r}\right)^k\,.
\end{multline}
The problem then reduces to find the coefficients $\vec{a}_k$ and the spectrum of the complex frequencies $\omega$, and it can be treated numerically with Leaver's CF method \cite{leaver_1, leaver_2} (see e.g. Ref.~\cite{pani_review} for more details on how to apply this method for a system of coupled differential equations).
\subsubsection{Classification of the spectrum}
From Eqs.~\eqref{eq:axial:eom} and~\eqref{eq:polar:eom} we can divide the QNM frequencies in five families, by looking at their limit when $v\to 0$. In the axial case they can be divided in axial gravitational modes $\omega_{AG,l}$ (corresponding to the Regge-Wheeler modes when $v\to 0$) and axial EM modes $\omega_{AE,l}$ (corresponding to the Maxwell modes). Similarly, in the polar case we divide them in polar gravitational modes $\omega_{PG,l}$ (corresponding to the Zerilli modes), polar EM modes $\omega_{PE,l}$ and polar scalar modes $\omega_{PS,l}$ (corresponding to the massless Klein-Gordon modes). We omitted an additional subscript for the overtone $n$, because we will restrict our analysis to the fundamental tone $n=0$ only.

Some results on the expected behavior of the QNMs have already been discussed in the literature so let us summarize them here. Ref.~\cite{lehner_2} numerically simulated the collision and ringdown of static, weakly charged ($Q\sim10^{-3}$) BHs in EMD for a wide range of $\eta$. The waveform extraction analysis was limited to the leading gravitational mode $l=2$ and scalar dipole mode $l=1$, finding that (i) for small charges the frequency does not depend sensitively on $\eta$ and (ii) there is no ISO-breaking for the gravitational modes within the numerical error. Ref.\,\cite{ferrari_1}, restricting to $\eta=1$, showed that the presence of the dilaton actually induces some ISO-breaking and that, as should expected, the deviation increases with the charge $v$. In Ref.~\cite{Konoplya:2001ji} the QNMs were computed for generic $\eta$ but only for the axial sector, finding that the frequency spectrum depends weakly on the dilaton coupling. These results show that ISO-breaking is not so pronounced in the gravitational modes, while it appears more evidently in the EM sector, although for the small electric charges considered in Ref.~\cite{lehner_2} the ISO-breaking is below the numerical error.

As we show below, our results agree with the conclusions of \cite{lehner_2,ferrari_1,Konoplya:2001ji} when the parameter space overlaps. However, we are able, for the first time, to monitor how the properties of the QNMs vary with both $v$ and $\eta$ for both the polar and axial sectors. The most important outcome is that, while it is true that gravitational modes do not allow to distinguish appreciably between different values of $\eta$, the dilaton coupling has a clear impact on the EM modes, in particular on the polar QNMs. Indeed we show that the degree of the EM ISO-breaking varies considerably with $\eta$, thus furnishing a clear signature of the dilaton coupling.
\begin{figure*}[htb]
\centering
\includegraphics[width=0.42\textwidth]{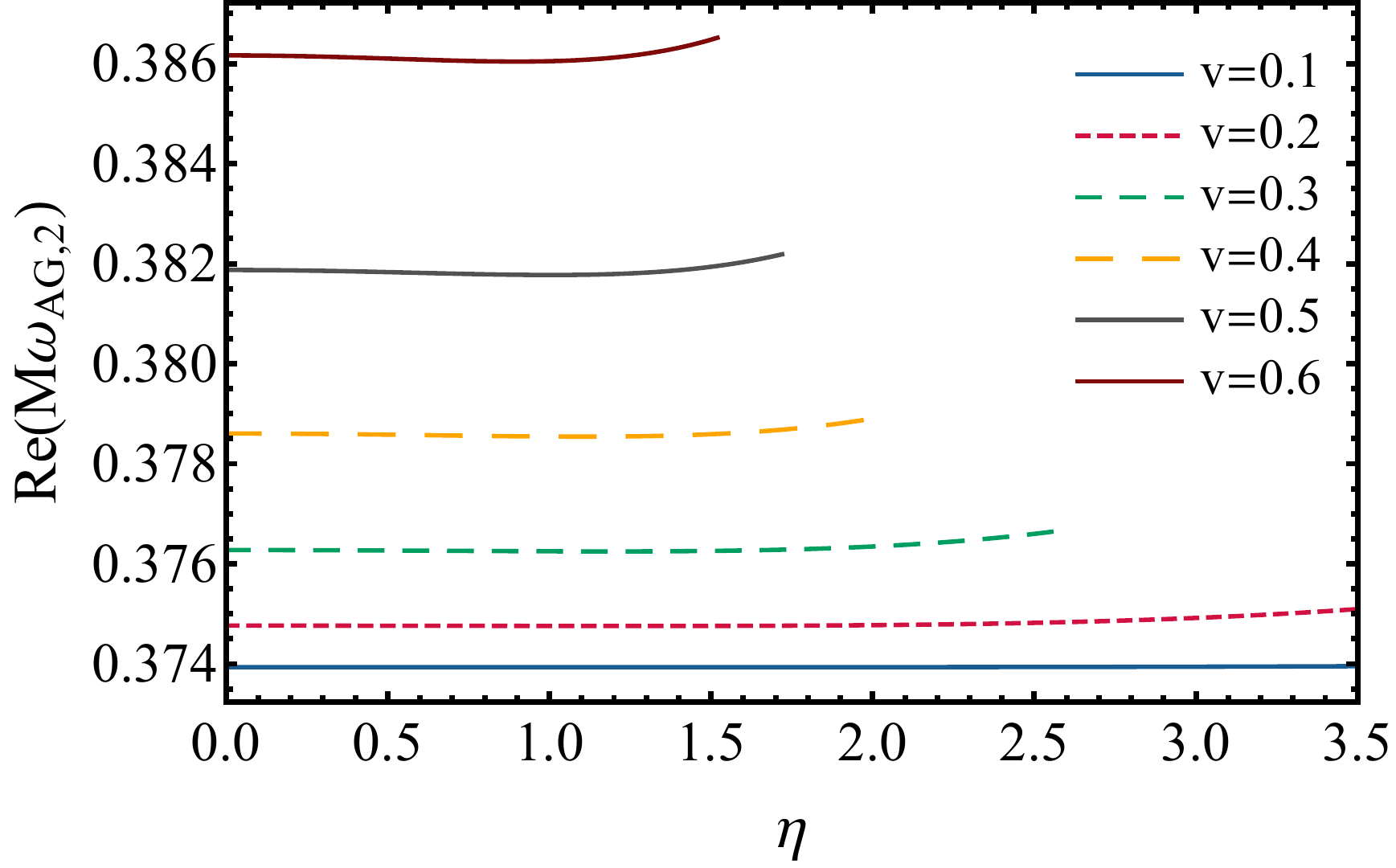}\,
\includegraphics[width=0.44\textwidth]{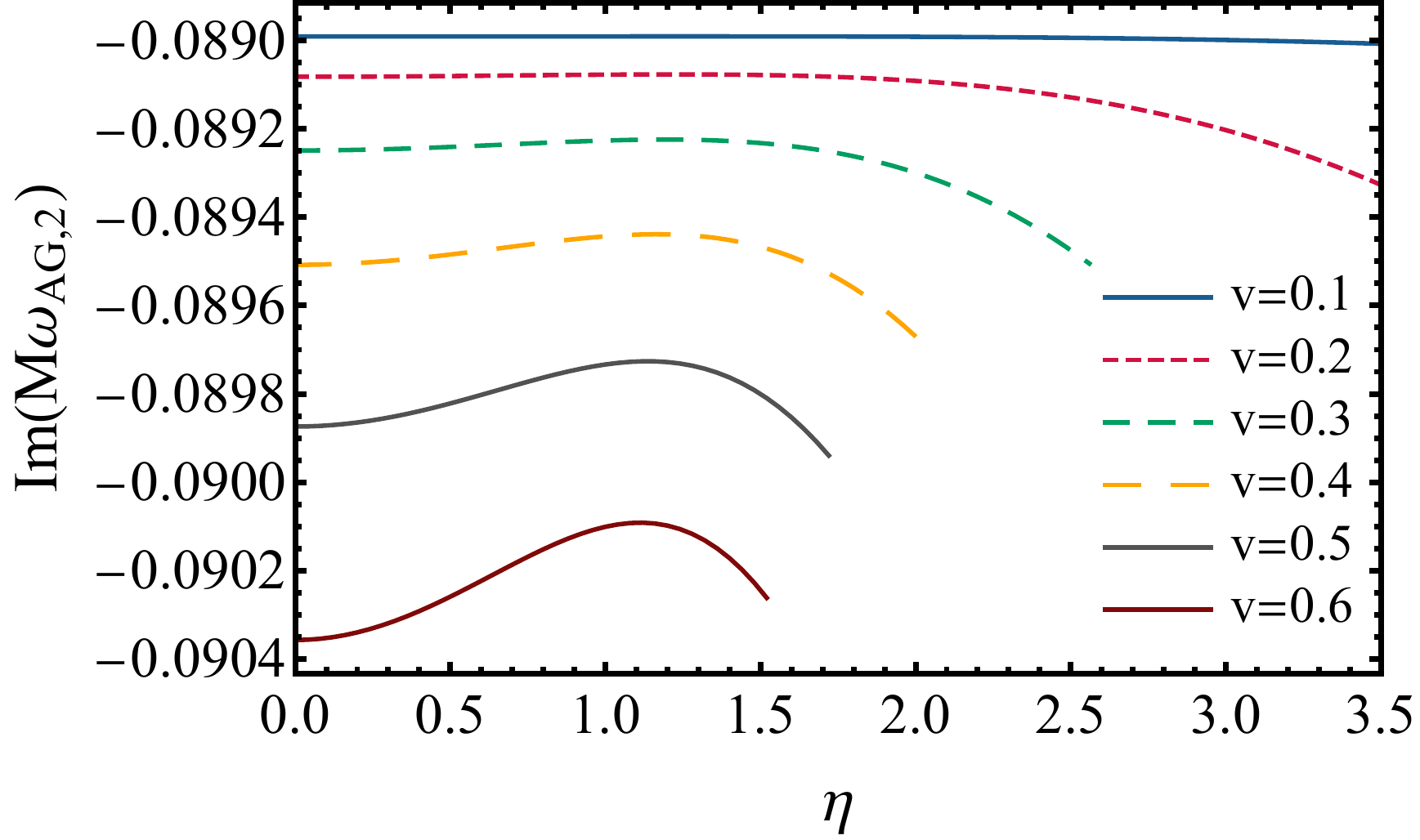}\\
\includegraphics[width=0.42\textwidth]{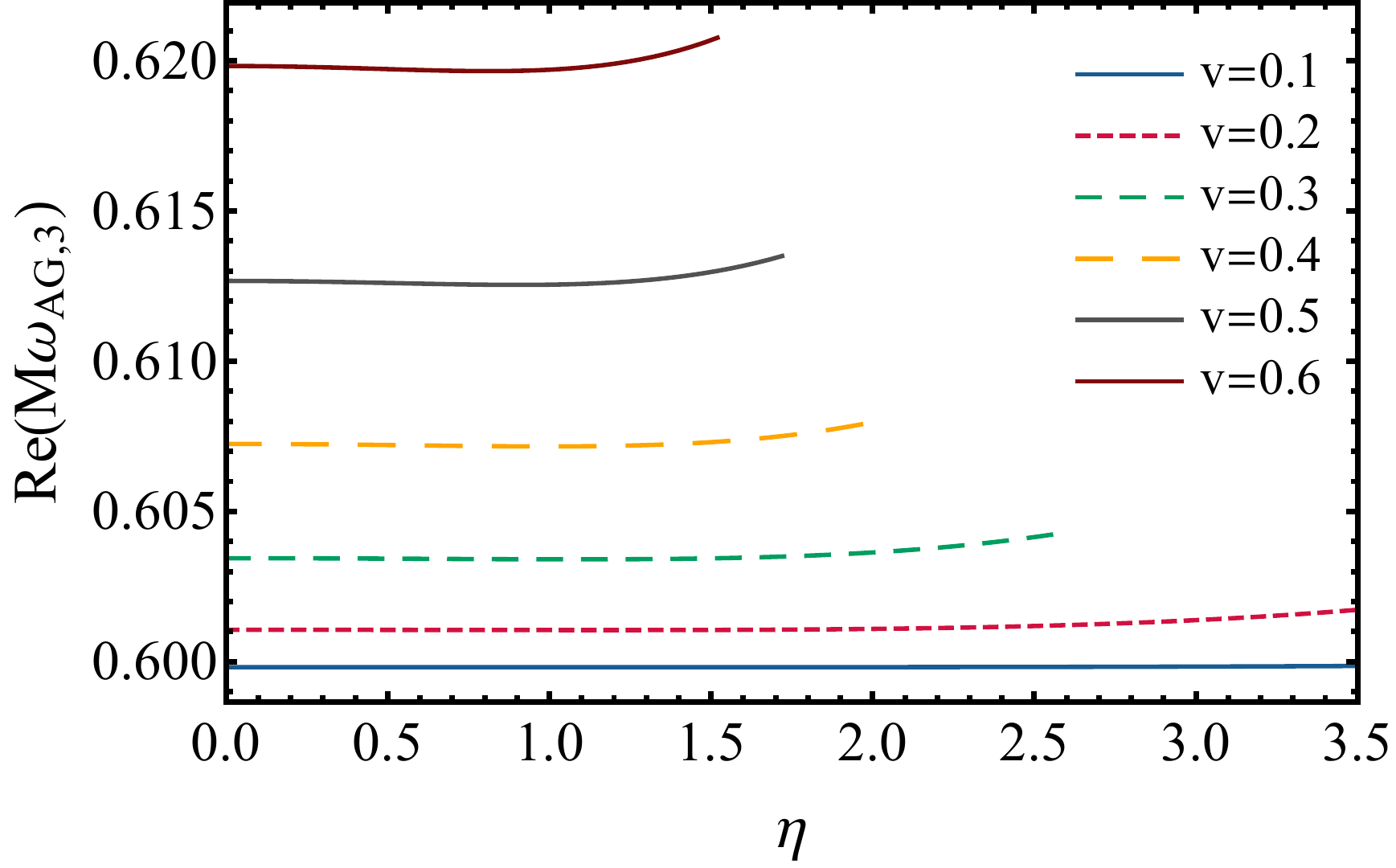}\,
\includegraphics[width=0.45\textwidth]{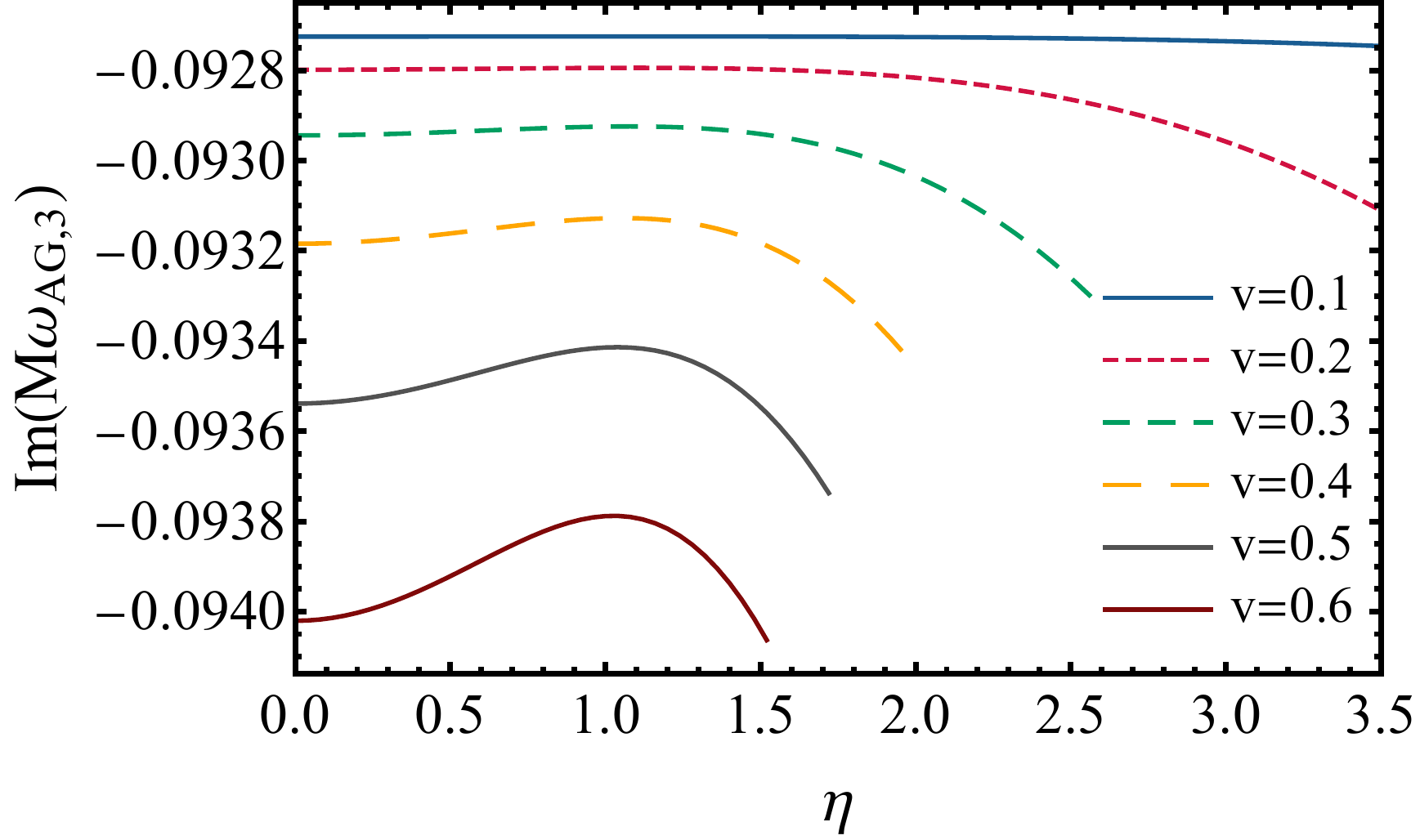}
\caption{Real (left) and imaginary (right) parts of the axial gravitational QNMs, $M\omega_{AG,l}$,  as a function of the dilaton coupling $\eta$, for $l=2$ (top) and $l=3$ (bottom) for different BH's charge-to-mass ratios $v$, computed using the $\mathcal{O}(v^2)$ equations.}
\label{plot:wg2}
\end{figure*}
%
\subsubsection{The gravitational modes}   
\label{sec:qnm:gr}
The gravitational modes exist for $l\geq2$. In the limit $v\to 0$, the least-damped modes are the $l=2$ and $l=3$ fundamental tones, given respectively by $M\omega_{AG,2}=M\omega_{PG,2}=0.3737-i\,0.0890$ and $M\omega_{AG,3}=M\omega_{PG,3}=0.5994-i\,0.0927$~\cite{Berti:2009kk}. In Fig.~\ref{plot:wg2} we show the real and imaginary parts of the axial QNM frequencies for $l=2$ and $l=3$ as a function of the dilaton coupling $\eta$, for different values of $v$. One can see that the frequencies depend very weakly on $eta$: the relative difference between the maximum and the minimum for the curves shown in Fig.~\ref{plot:wg2} remains below $0.6 \%$, for both real and imaginary parts.

\begin{figure*}[htb]
\centering
\includegraphics[width=0.42\textwidth]{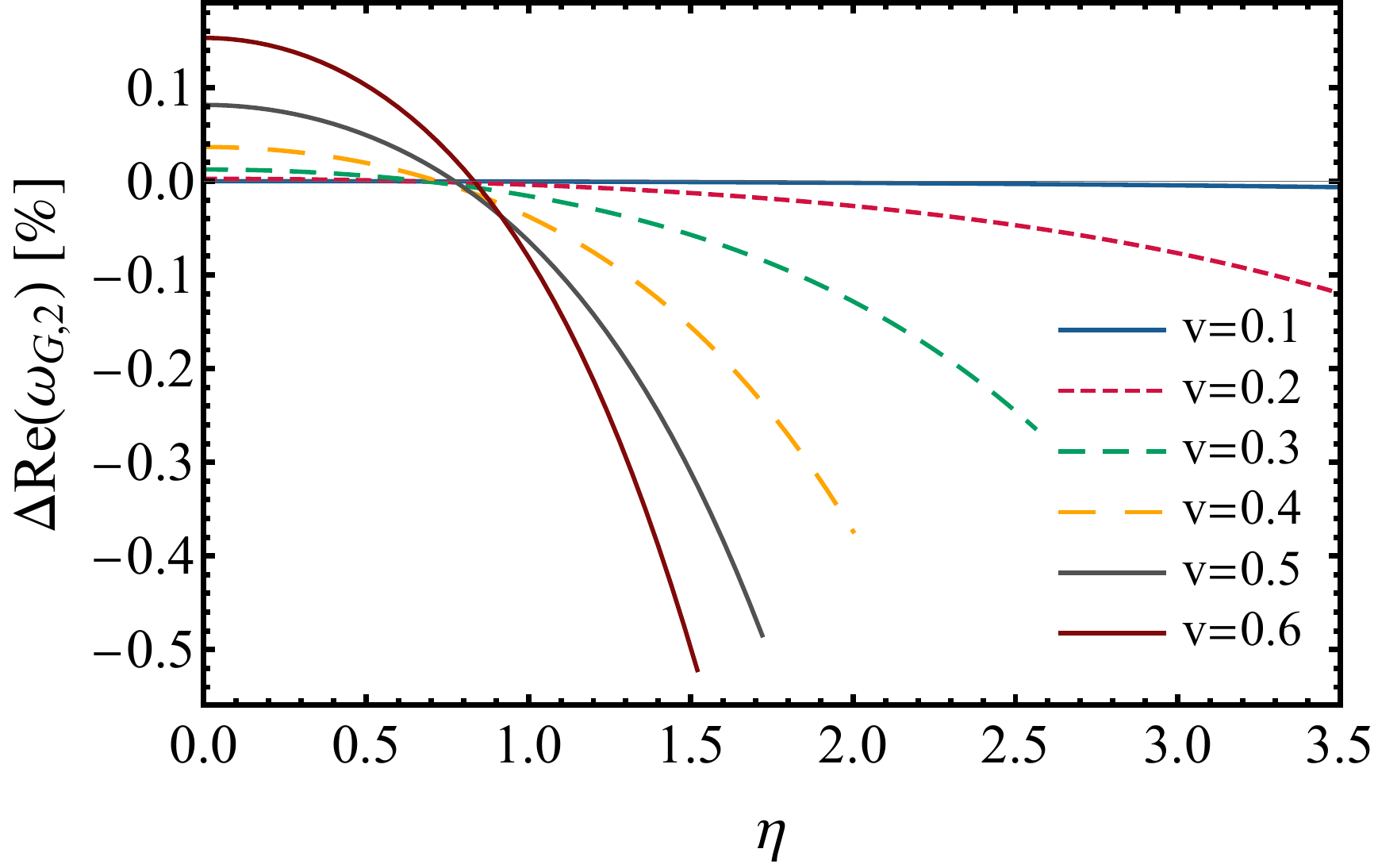}\,
\includegraphics[width=0.41\textwidth]{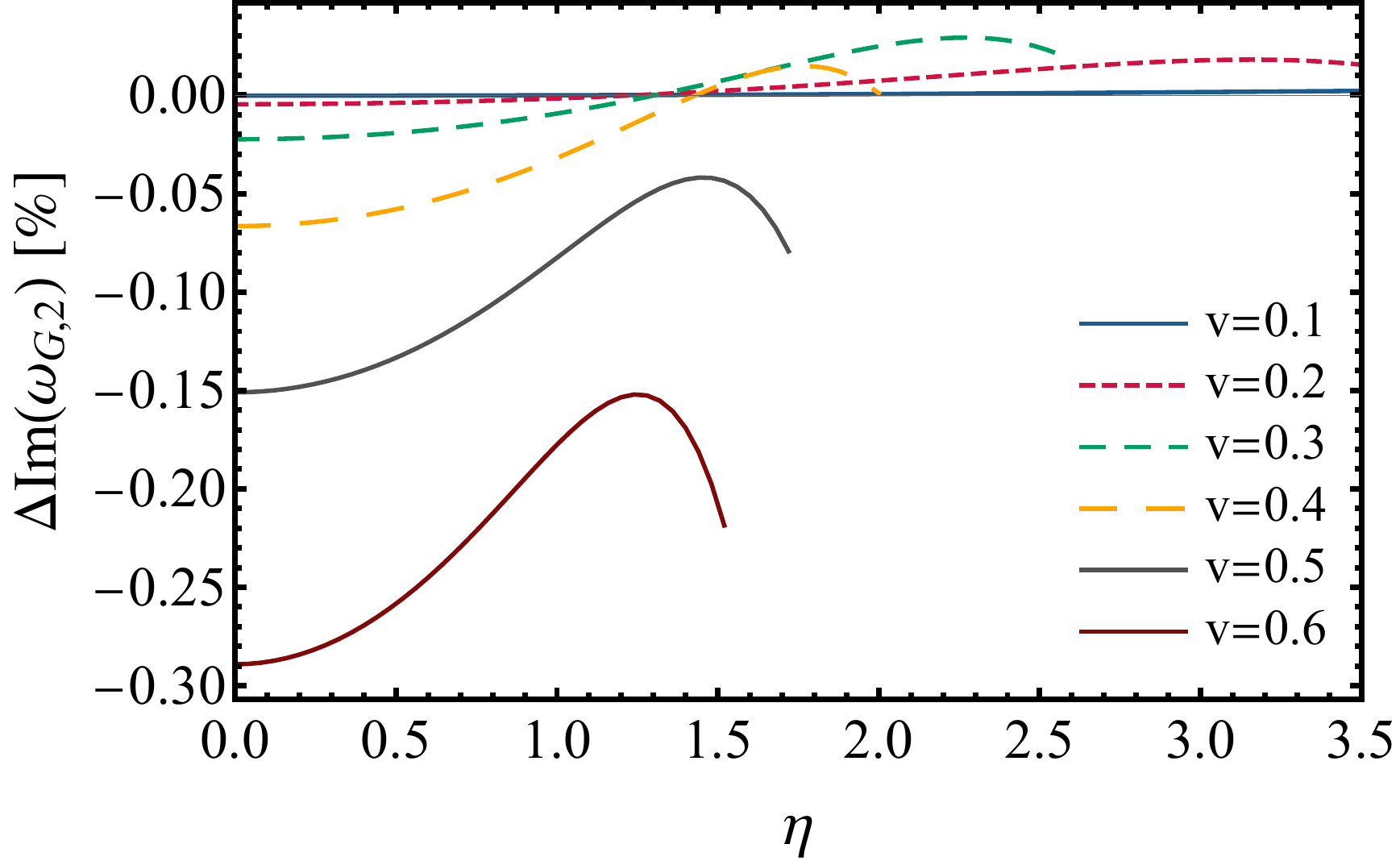}\\
\includegraphics[width=0.42\textwidth]{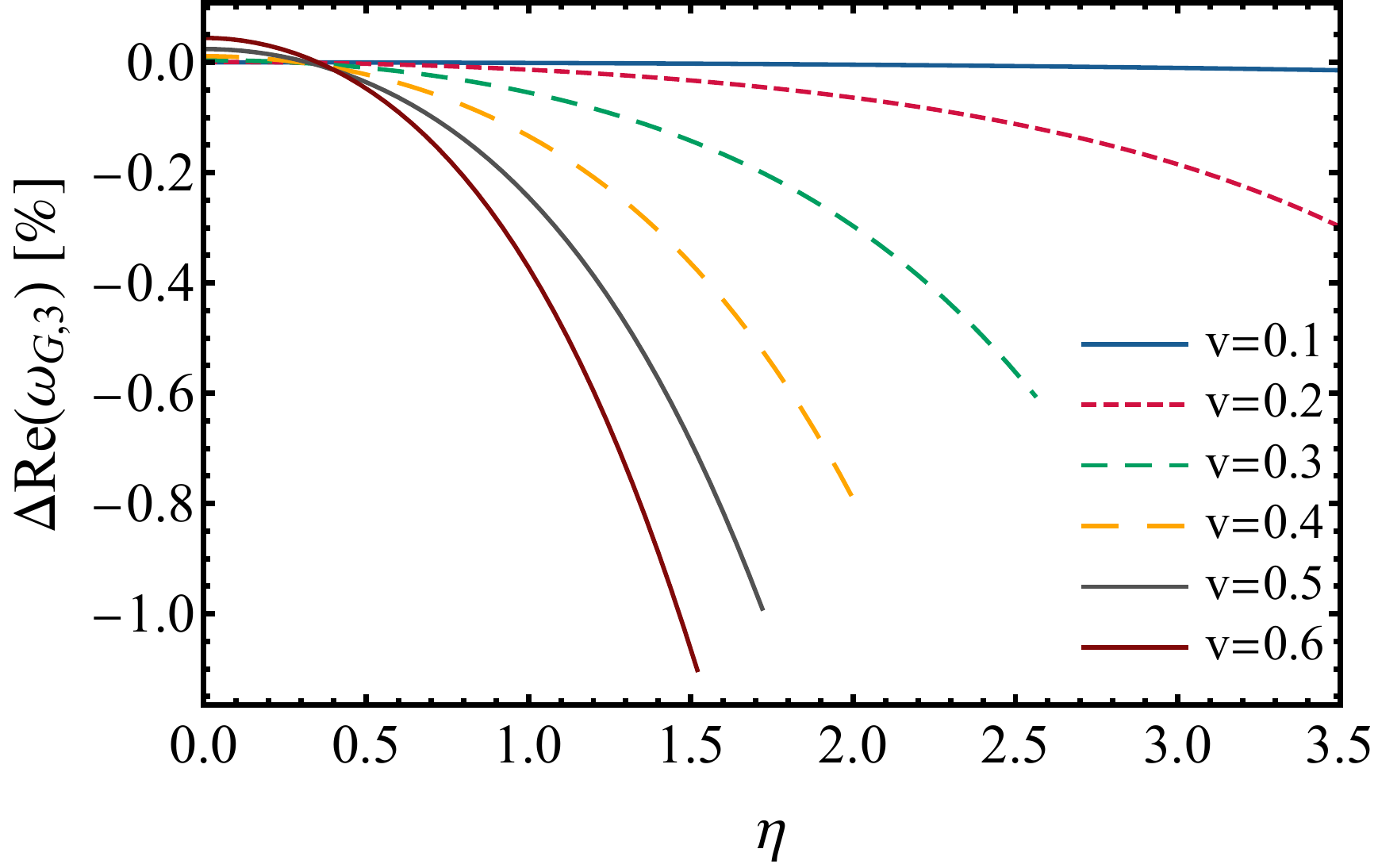}\,
\includegraphics[width=0.42\textwidth]{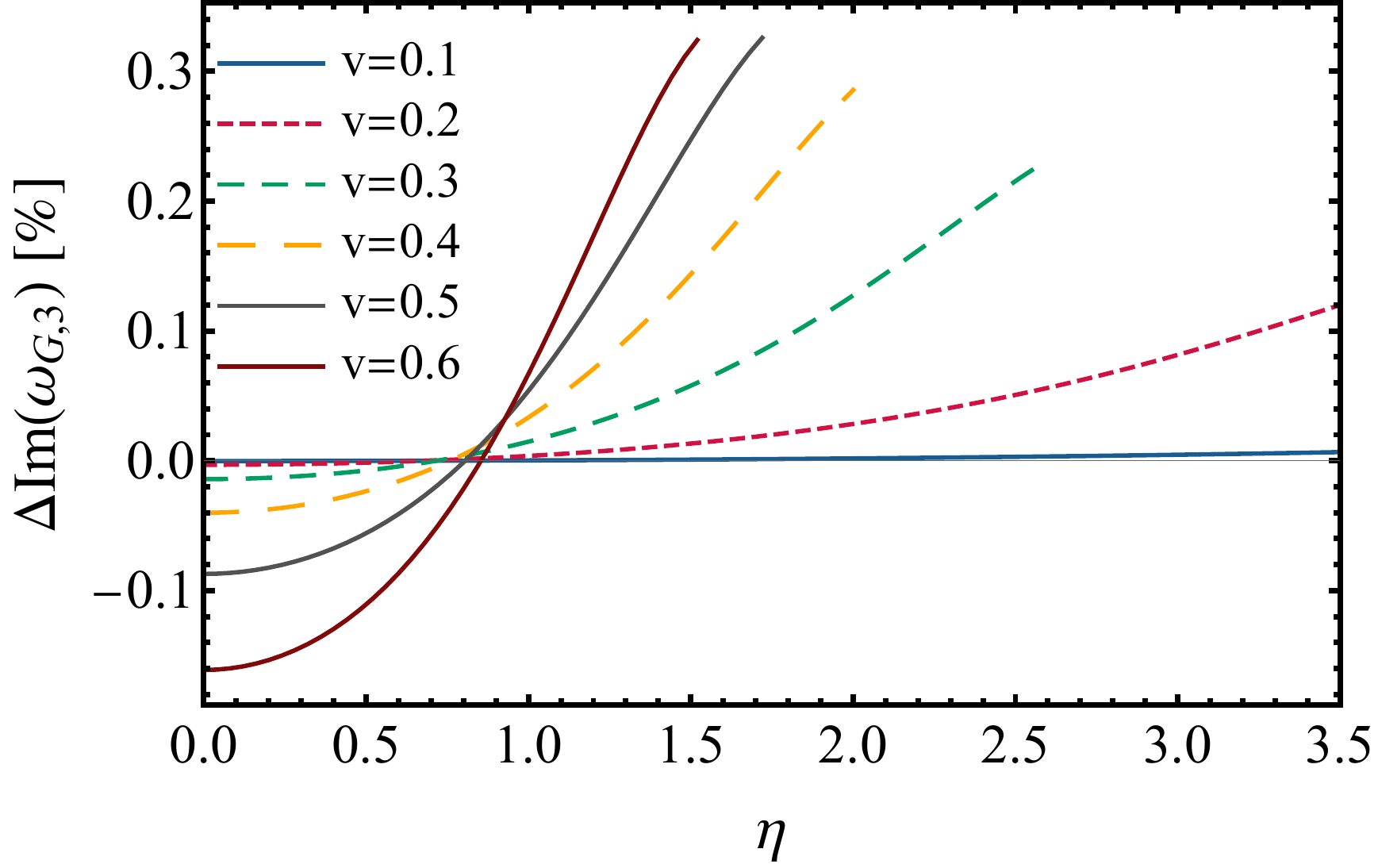}
\caption{ISO-breaking [cf. Eq.~\eqref{eq:delta:iso}] between the polar and axial gravitational QNMs for $l=2$ (top) and $l=3$ (bottom).}
\label{plot:delta:g}
\end{figure*}

The corresponding polar modes are analogous qualitatively and quantitatively, as we show in Fig.~\ref{plot:delta:g} where we plot the relative percentage difference
\be
\label{eq:delta:iso}
\Delta\text{Re}\left(\omega_{G,l}\right)=100\times\frac{\text{Re}\left(\omega_{PG,l}\right)-\text{Re}\left(\omega_{AG,l}\right)}{\abs{\text{Re}\left(\omega_{AG,l}\right)}}
\ee
and similarly for the imaginary part. We see that ISO-breaking grows only moderately with $v$, being almost negligible at small $v\leq 0.2$ and becoming of the order of the percent for $v=0.6$. Therefore we observe a very weak ISO-breaking in the gravitational sector, in agreement with~\cite{ferrari_1,lehner_2}.

\paragraph*{Approximation error.}
It is useful to estimate the percentage error made in working with the $\mathcal{O}(v^2)$ approximation. To this aim, we derived the exact form of the axial perturbed EOM at all orders in $v$ and we compared the $l=2$ axial gravitational modes with the ones from the $\mathcal{O}(v^2)$ EOM (see Appendix \ref{appendix:2}). We find that the error remains at the order of $\sim 1\%$ in the parameter space we consider (cf. Fig.~\ref{plot:delta:full}). In particular, we note that this error is comparable to the amount of ISO-breaking that we find in the gravitational sector.

\subsubsection{The electromagnetic modes}
\label{sec:qnm:em}
The EM modes exist for $l\geq1$. As shown in Refs.~\cite{collision_1,collision_2,Liebling:2016orx,Cardoso:2016olt}, these modes can become significant for the radiation emitted by the merger of charged BHs. In particular, Refs.~\cite{collision_1,collision_2} numerically studied head-on BH collisions in Einstein-Maxwell theory ($\eta=0$) for equal \cite{collision_1} and opposite \cite{collision_2} charge-to-mass ratio, while Ref.~\cite{Liebling:2016orx} simulated the inspiral of weakly charged Reissner-Nordstr\"{o}m BHs for different initial configurations. A generic prediction of these studies is that the process is always accompanied by the emission of both EM waves and GWs, with the ringdown part being described by a superposition of both EM and GW QNM frequencies. 
In addition, for the head-on collisions, it was shown that while for equal charges the EM wave emission is always subdominant with respect to GWs~\cite{collision_1}, for opposite charges, $l=1$ EM waves become the dominant channel of radiation emission already for moderate values of $\abs{v}\geq 0.37$~\cite{collision_2}. Therefore, depending on the initial binary parameters, EM wave emission and EM QNMs can constitute a non-negligible part of the radiation and their study can be relevant for the purposes of BH spectroscopy.

\begin{figure*}[htb]
\centering
\includegraphics[width=0.42\textwidth]{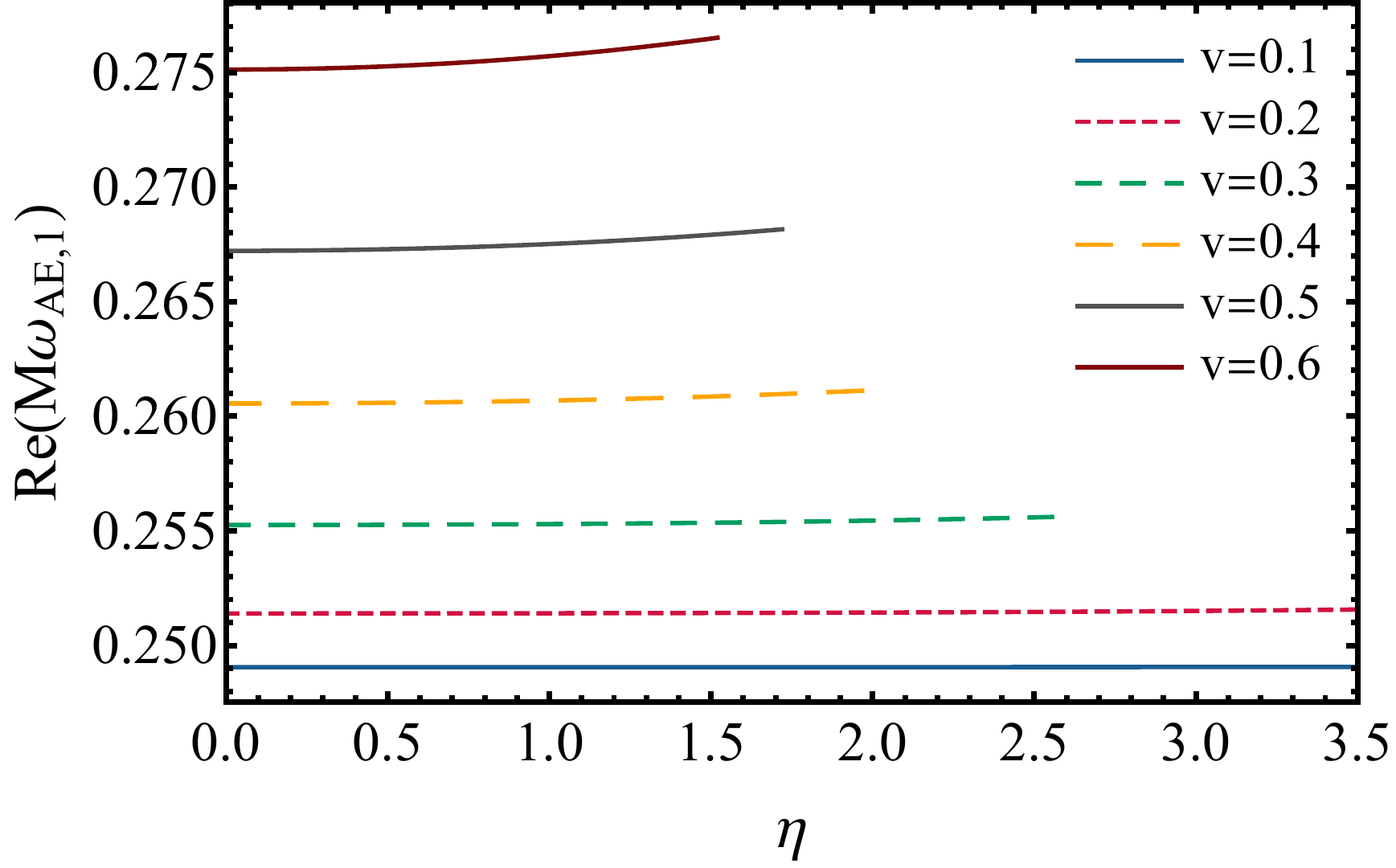}\,\hspace{5mm}
\includegraphics[width=0.42\textwidth]{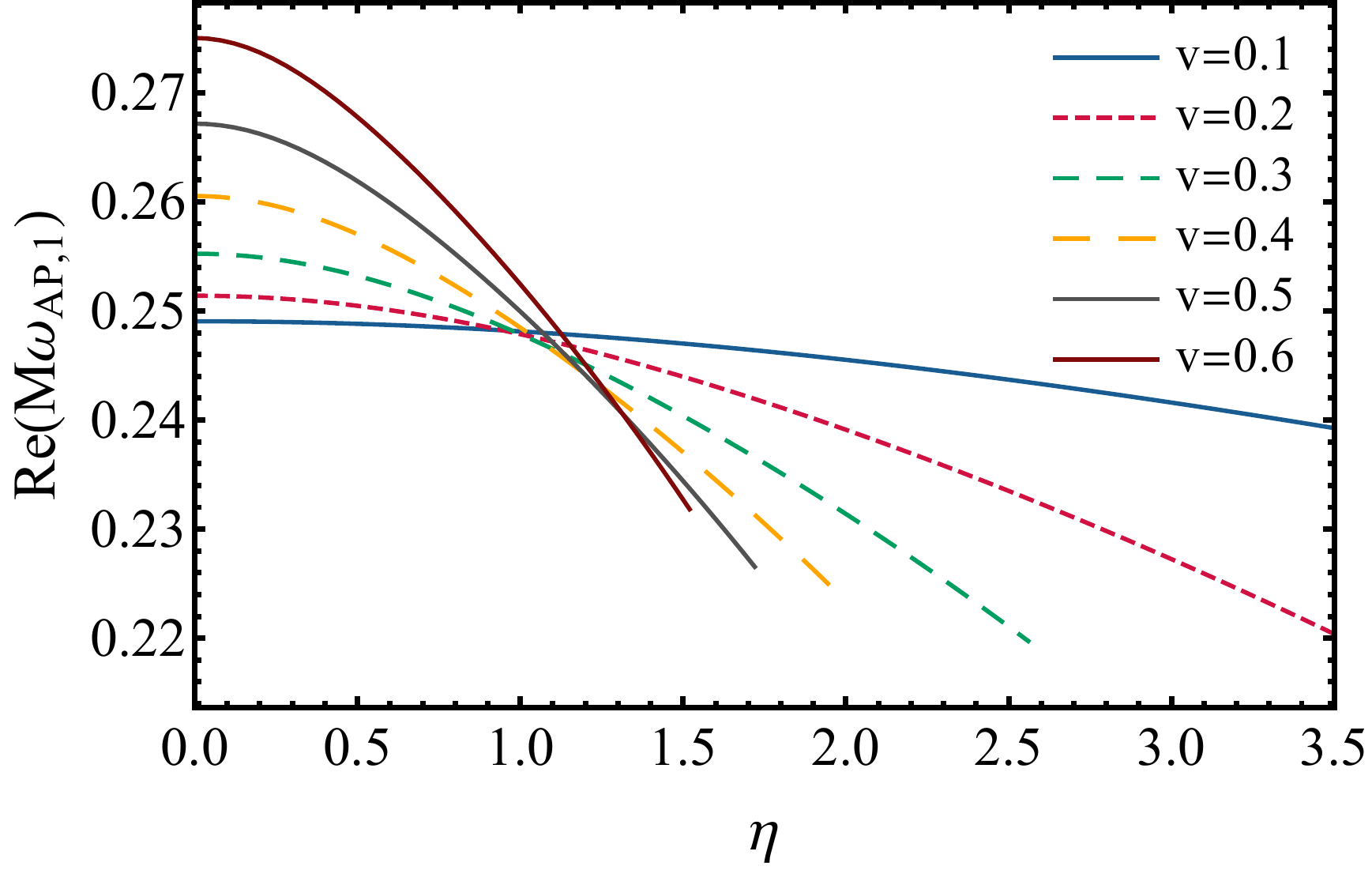}\\
\includegraphics[width=0.44\textwidth]{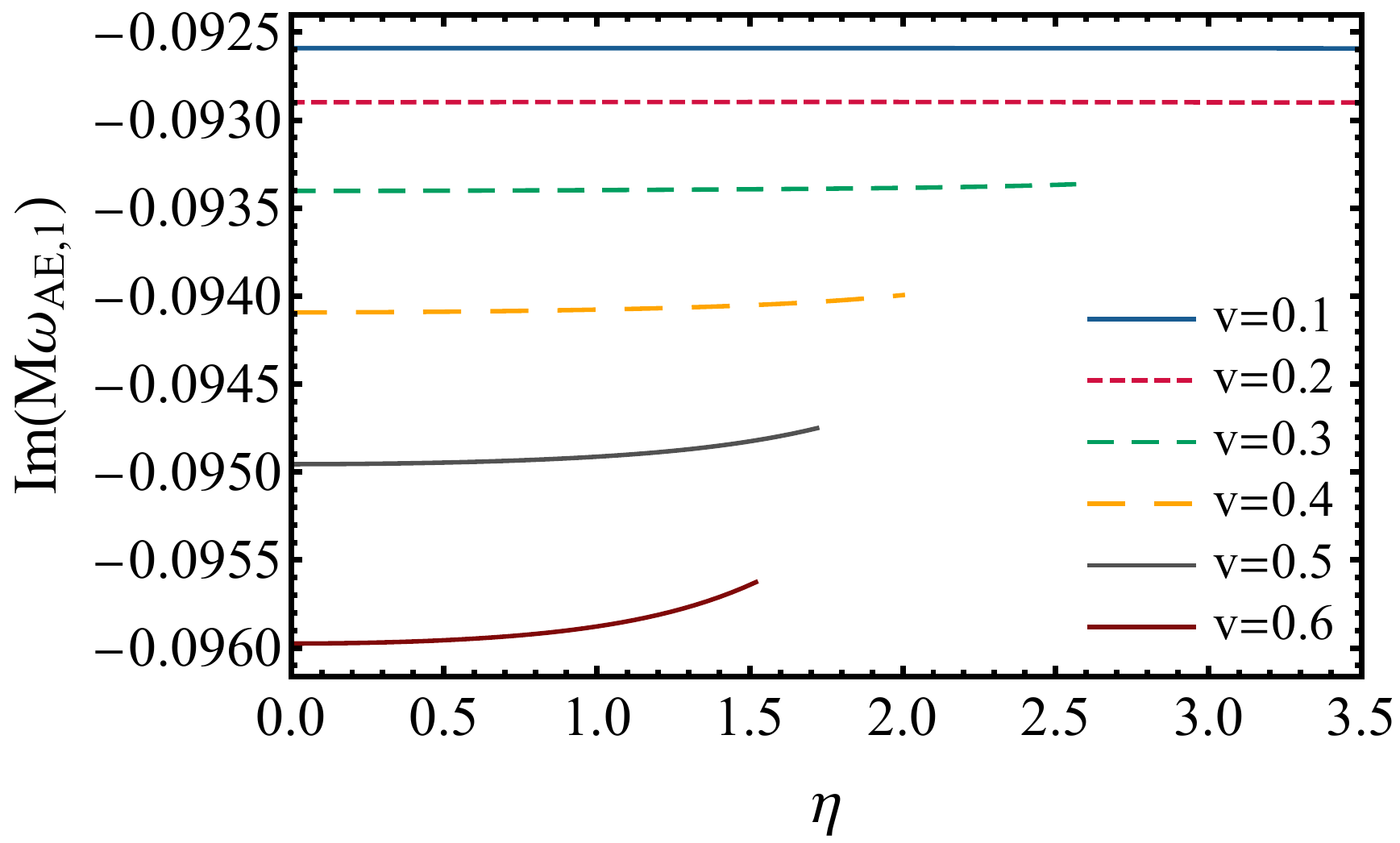}\,
\includegraphics[width=0.44\textwidth]{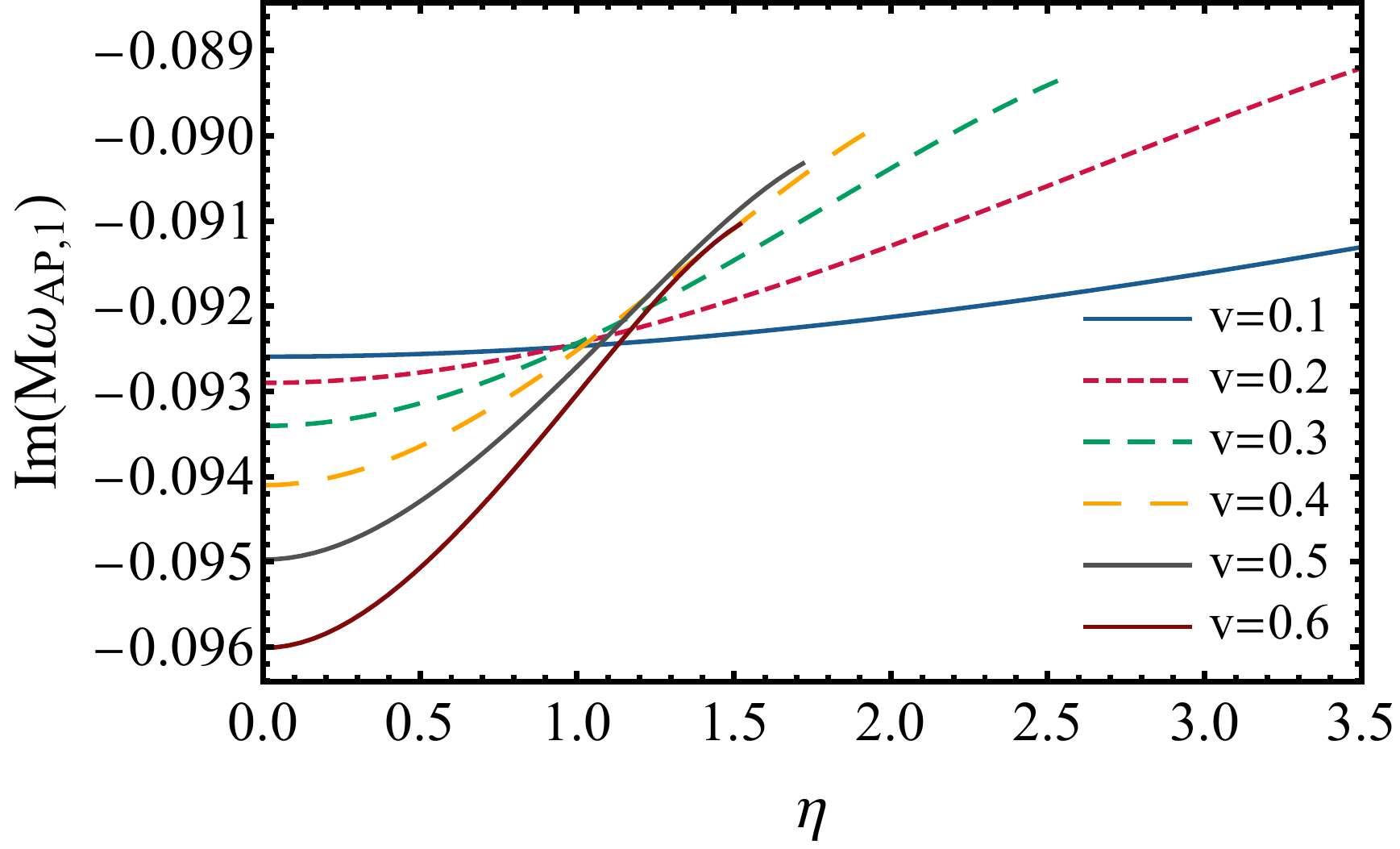}
\caption{Real (top) and imaginary (bottom) parts of the EM axial QNMs (left), $M\omega_{AE,l}$, and polar QNMs (right), $M\omega_{AP,l}$, for $l=1$, computed using the $\mathcal{O}(v^2)$ equations.}
\label{plot:we1}
\end{figure*}

\begin{figure*}[htb]
\centering
\includegraphics[width=0.42\textwidth]{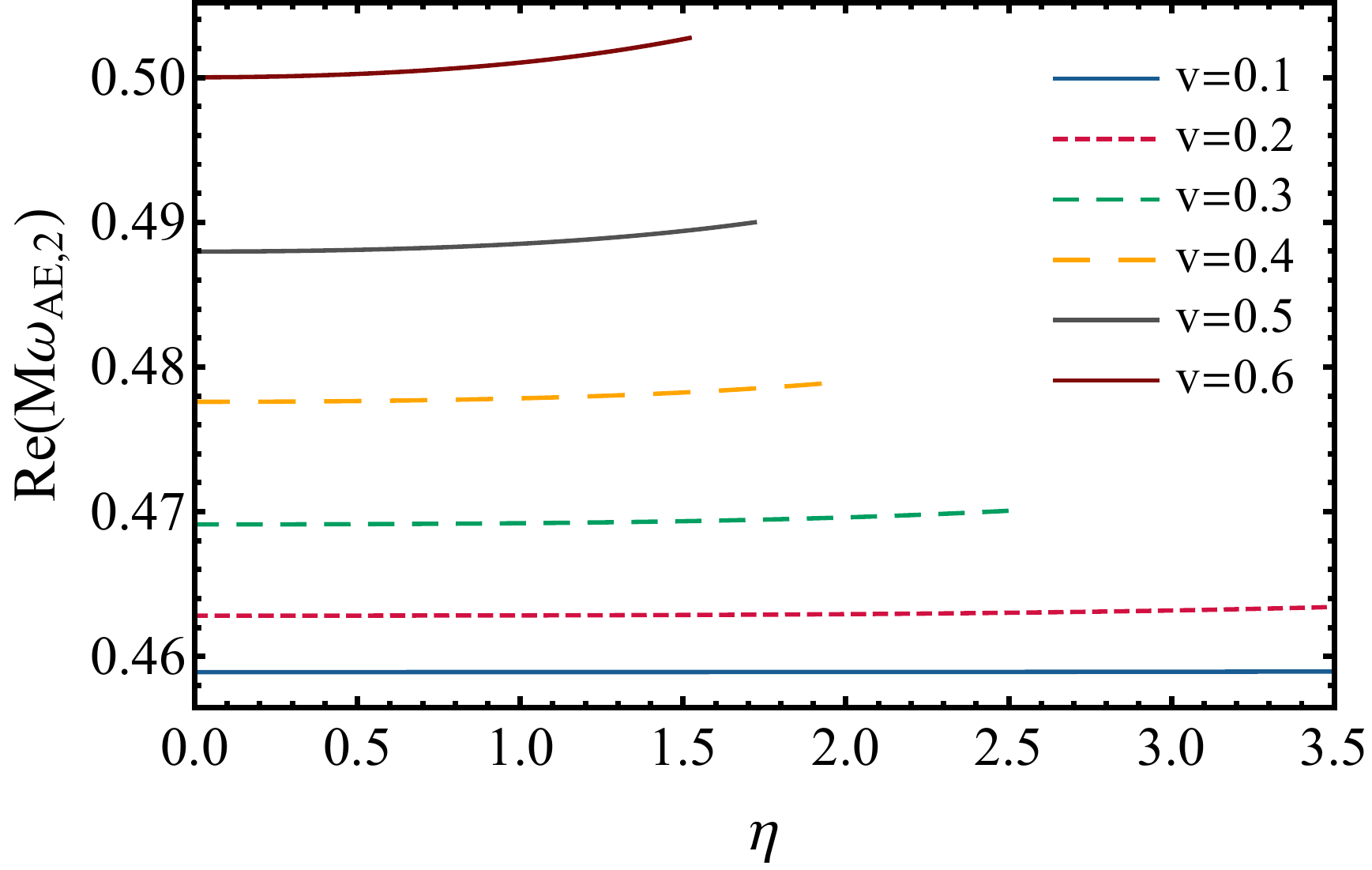}\,\hspace{5mm}
\includegraphics[width=0.42\textwidth]{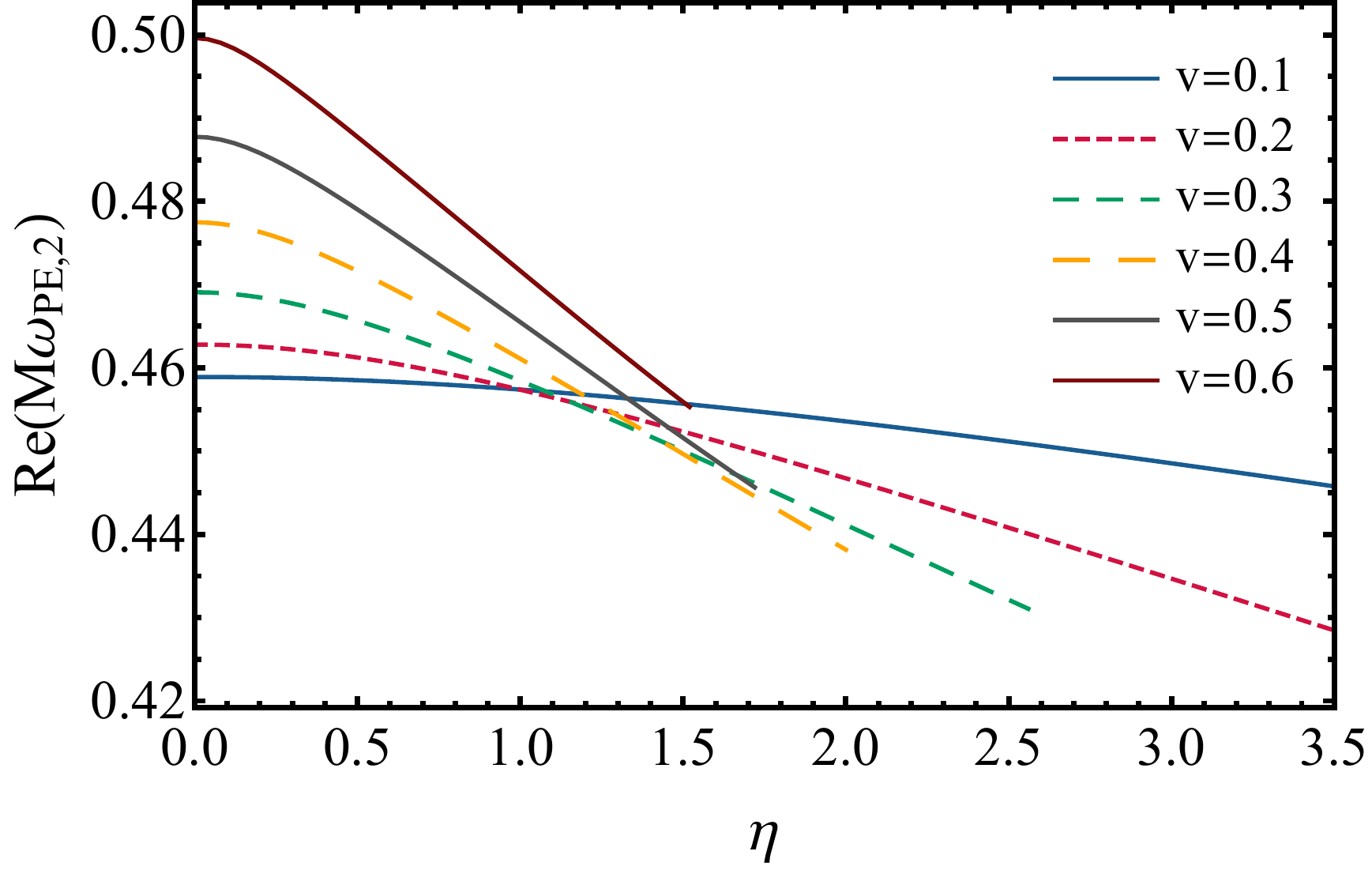}\\
\includegraphics[width=0.44\textwidth]{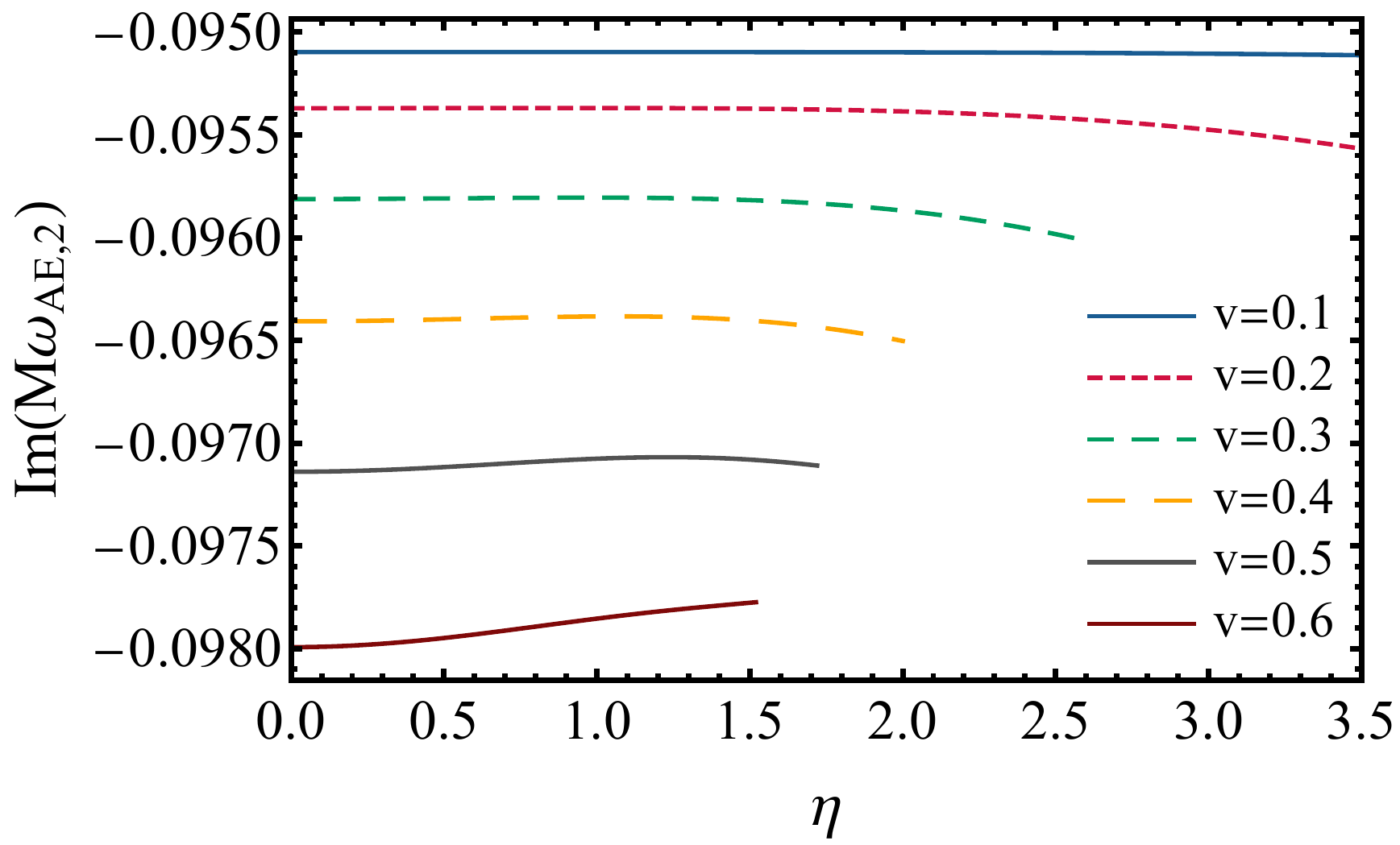}\,
\includegraphics[width=0.44\textwidth]{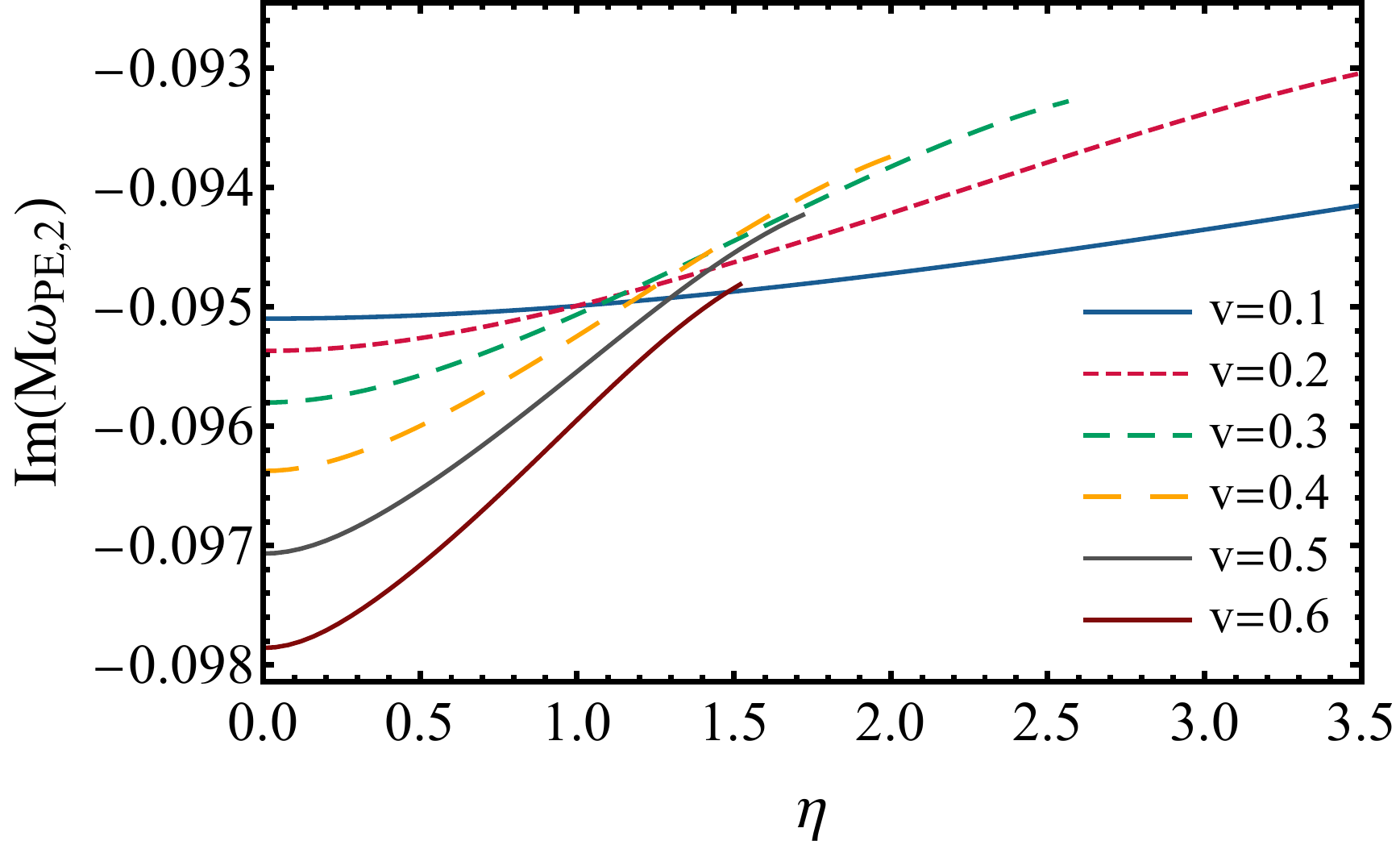}
\caption{Same as Fig.~\ref{plot:we1} but for $l=2$.}
\label{plot:we2}
\end{figure*}

For concreteness, we focus on the axial and polar EM modes for $l=1$ and $l=2$. Our results, for $0.1\leq v\leq 0.6$, are shown in Fig.~\ref{plot:we1} (for $l=1$) and Fig.~\ref{plot:we2} (for $l=2$). In the limit $v\to 0$, these QNMs coincide with the fundamental EM modes on a Schwarzschild background, i.e. $M\omega_{AE,1}=M\omega_{PE,1}=0.2483-i\,0.0925$ and $M\omega_{AE,2}=M\omega_{PE,2}=0.4576-i\,0.0950$~\cite{Berti:2009kk}. As can be easily seen already at a qualitative level, there is a marked difference between the axial and the polar modes for sufficiently high $\eta$. In particular, the polar QNMs have a much stronger dependence on $\eta$ which can be understood from the fact that the dilaton only couples directly to the polar EOM.


\begin{figure*}[htb]
\centering
\includegraphics[width=0.42\textwidth]{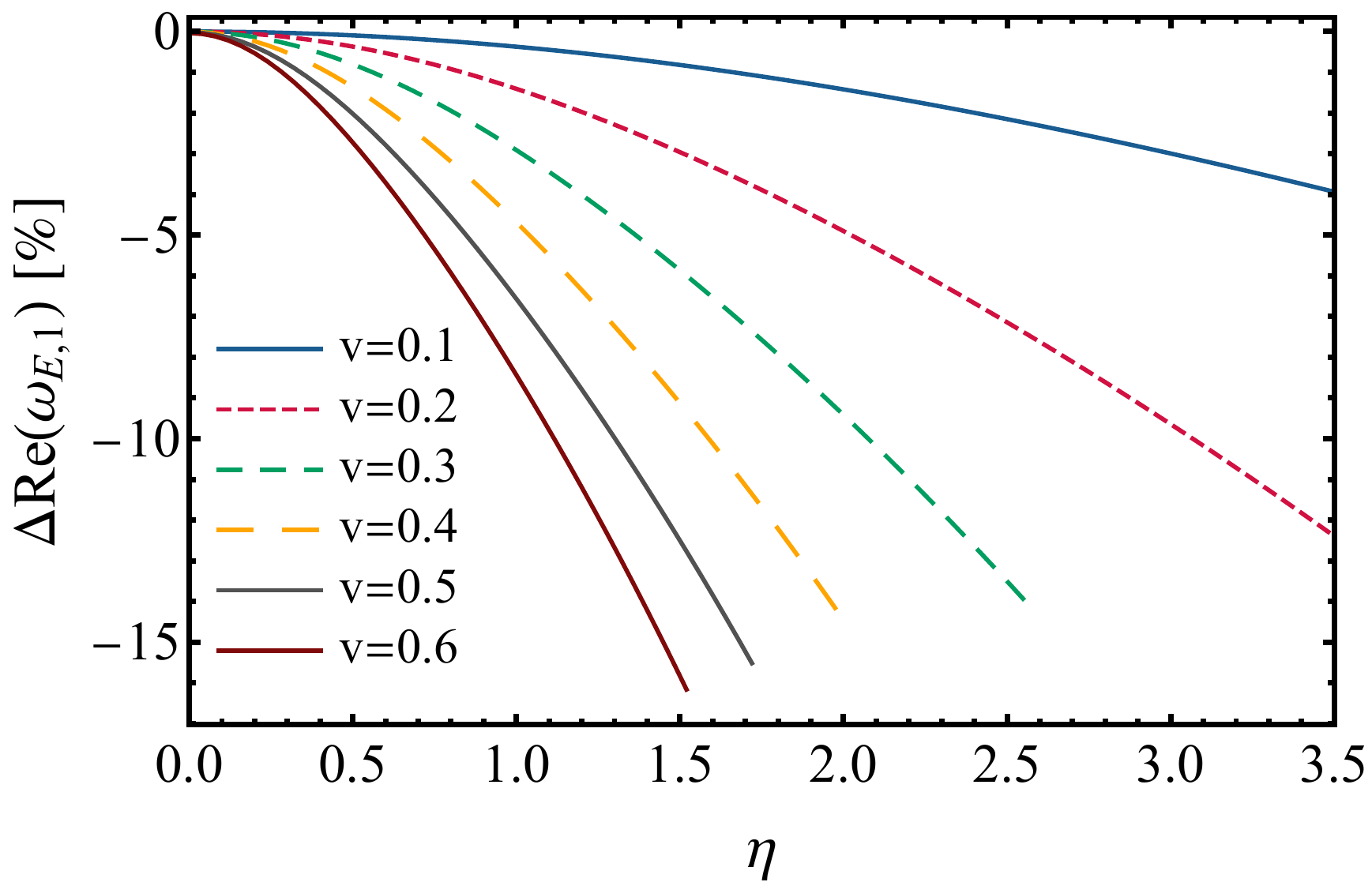}\,
\includegraphics[width=0.41\textwidth]{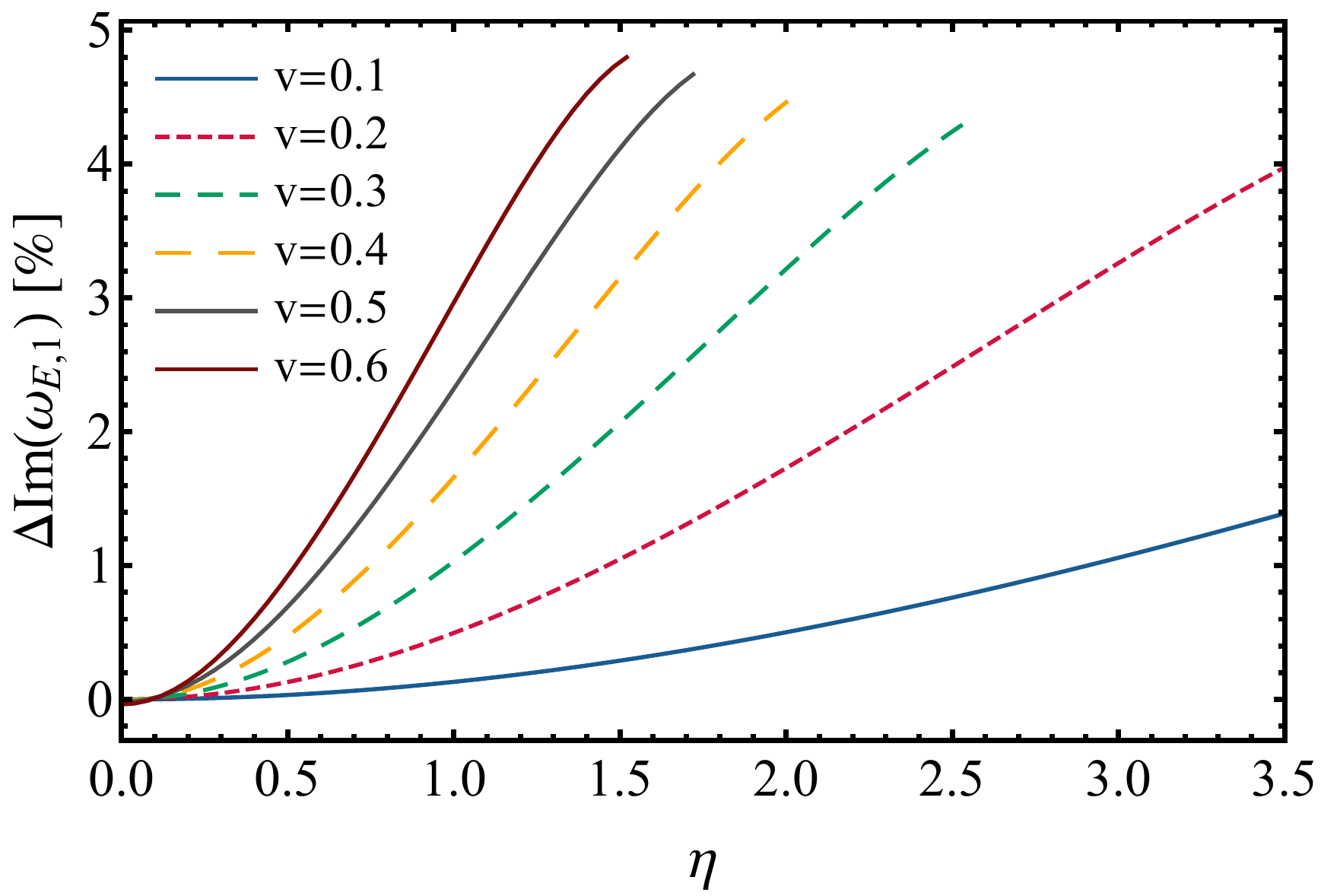}\\
\includegraphics[width=0.42\textwidth]{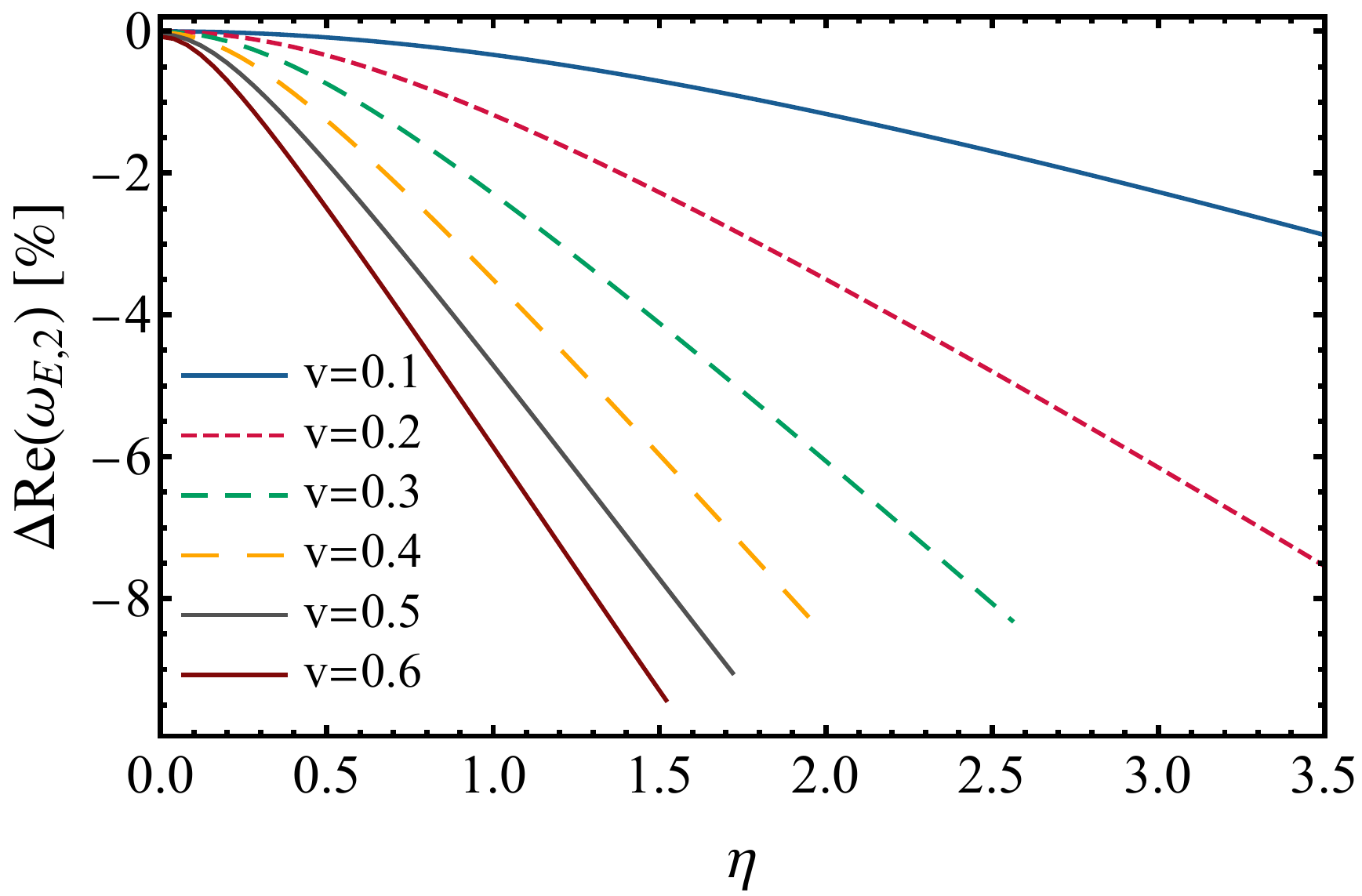}\,
\includegraphics[width=0.42\textwidth]{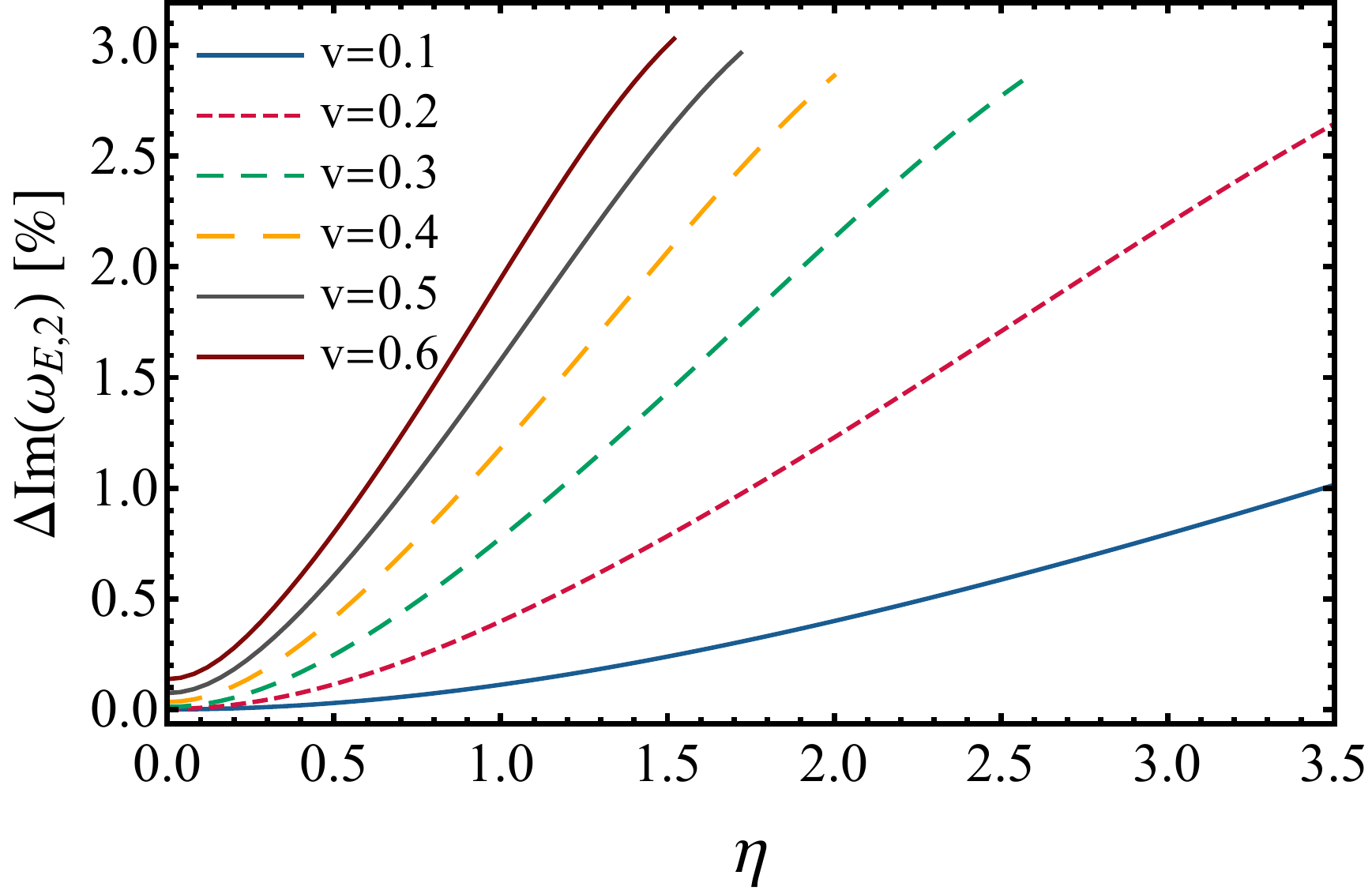}
\caption{ISO-breaking [cf. Eq.~\eqref{eq:delta:iso}] of the EM QNMs for $l=1$ (top) and $l=2$ (bottom).}
\label{plot:delta:e}
\end{figure*}

This difference is more easily seen in Fig.~\ref{plot:delta:e} where we show the percentage of ISO-breaking, evaluated as in Eq.~\eqref{eq:delta:iso}. The difference between polar and axial modes is very small for $\eta\sim 0$ but grows monotonically with $\eta$ and $v$.\footnote{We note that the difference should be exactly zero for $\eta=0$ because of the known isospectrality of the Reissner-Nordstr\"{o}m QNMs~\cite{chandra_book}. The very small departure from zero at large $v$ can be ascribed to the small charge approximation that we employed.} Isospectrality of the real part of the polar and axial frequencies is broken up to $\sim 15 \%$ for $l=1$ and $\sim 8\%$ for $l=2$, while for the imaginary part the effect is smaller, but still more pronounced than in the gravitational sector. Therefore ISO-breaking in the EM sector provides a clear signature to distinguish EMD BHs in the $(v,\eta)$ plane. In fact, in Sec.~\ref{sec:df} we shall see that the EM ISO-breaking is rooted in the coupling between the vector field and the dilaton.

\subsubsection{The scalar modes}
Unless $\eta=0$, the dilaton perturbations couple dynamically to the other fields, therefore inducing the presence of scalar modes. From the action \eqref{eq:action}, one expects that the importance of the scalar radiation grows with $\eta$, being almost negligible when $\eta\ll1$~\cite{lehner_2}. This is already visible in the above analysis of EM QNMs, where we saw that larger values of $\eta$ are also accompanied by an increasing of EM ISO-breaking. 

\begin{figure*}[htb]
\centering
\includegraphics[width=0.42\textwidth]{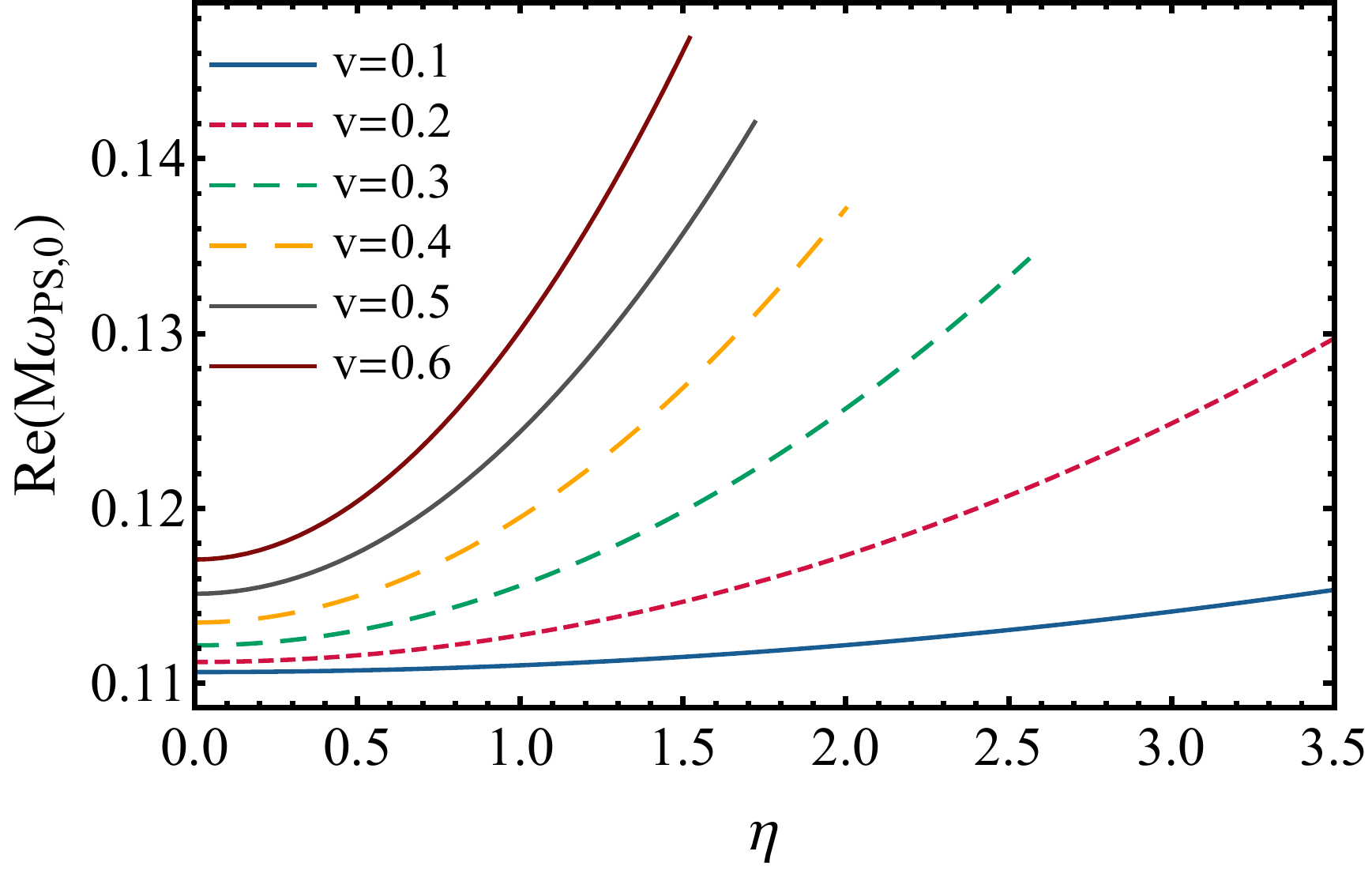}\,
\includegraphics[width=0.44\textwidth]{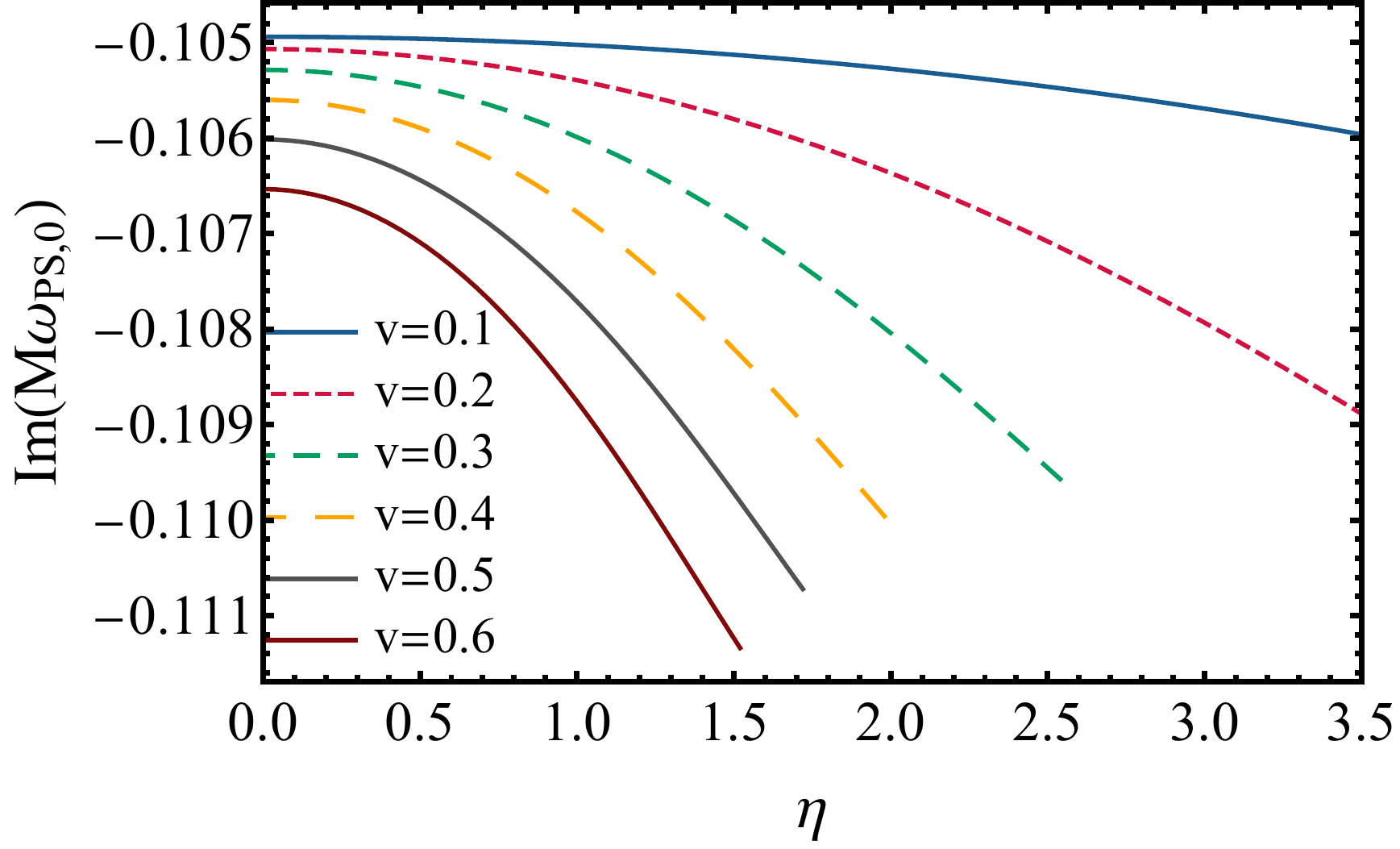}
\caption{Real (left) and imaginary (right) part of the scalar QNM, $M\omega_{PS,l}$, for $l=0$.}
\label{plot:ws}
\end{figure*}

A possible consequence of the presence of the dilaton is the possibility that it could induce instabilities in this BH spacetime. In fact, it was argued in Ref.~\cite{lehner_2} that the presence of the dilaton could induce tachyonic-like instabilities for sufficiently large coupling constant $\eta$. We did not find any evidence for an instability when computing the scalar QNMs. In particular, in Fig.~\ref{plot:ws} we show the scalar QNM for $l=0$, where it can be seen that the imaginary part is always negative, thus indicating that these modes always decay and are therefore stable (the same conclusion remains valid for $l=1$ and $l=2$). For reference, we note that the fundamental Klein-Gordon mode on a Schwarzschild background is given by $M\omega_{PS,0}=0.1105-i\,0.1049$.

Scanning the parameter space we were unable to find evidence for an unstable mode. In fact, by evaluating numerically the potential~\eqref{eq:v0} for generic values of $\eta$ and $v$, i.e. with no small-charge approximation (see Appendix~\ref{appendix:3}), we find that the potential is always positive definite for any value of $\eta$ and $v$. This fact is a proof that the $l=0$ modes do not suffer from instabilities~\cite{chandra_book}. We believe that this discrepancy might be due to the fact that for the values of $\eta$ for which Ref.~\cite{lehner_2} finds an instability, the small-charge approximation they employ is not valid, as we argued in Sec.~\ref{sec:formalism}.

In Appendix\,\ref{appendix:3} we clarify this point for a specific example: we consider the $l=0$ potential in the limit $\eta\to\infty$, without any restriction on the magnitude of $v$. In this limit, according to~\cite{lehner_2}, the spacetime should be unstable for arbitrarily small values of $v$. However, we find that the potential is always positive outside the event horizon. Moreover, we repeat the same analysis by perturbing the dilaton on the background~\eqref{eq:line:1}~--~\eqref{eq:matter:1} while keeping the vector field fixed, which is closer to the spirit of the calculation done in Ref.~\cite{lehner_2}. In this case we \emph{do} find the occurrence of an instability.

Although restricted to the limit $\eta\to\infty$, these results highlight what is probably a general lesson: when pushed beyond a consistent weak-field limit, a background analysis of the dilaton perturbations can lead to misleading results and must be validated against an exact treatment. More details are provided in Appendix~\ref{appendix:3}.
%

\subsection{The Dudley-Finley approximation}
\label{sec:df}
In Refs.~\cite{df_1,df_2} an approximate approach to compute the perturbation equations was introduced by Dudley and Finley, motivated by the difficulty of separating radial and angular perturbations in the Kerr-Newman spacetime. In the DF approximation the metric and the matter fields are perturbed separately. This method should be valid as long as the matter fields do not induce large deviations from vacuum GR, i.e. when the effects of matter are already weak at the background level. In the case of the Reissner-Nordstr\"{o}m black hole this expectation was confirmed in~\cite{berti_slow_1}, where the DF QNMs were found in good agreement with the exact ones for $v\lesssim 0.5$. It is reasonable to expect that a similar agreement remains valid in the more general case of EMD theory. 

The original DF method consists in perturbing each field independently from the others. We have seen that, while the gravitational modes are only weakly sensitive to the presence of the dilaton, EM modes are quite sensitive to the coupling to the dilaton. It is then reasonable to employ a modified DF scheme in which (i) the gravitational field is varied independently and (ii) the vector and scalar fields are varied together but independently from the metric.\footnote{Notice that, in the DF approach, metric or matter perturbations are turned off from the very beginning when one derives the perturbed EOM. An alternative approach could be to turn off the degrees of freedom at the end, once the EOM have already been obtained. For a discussion and a comparison of these two approaches see \cite{berti_df}.}

\begin{figure*}[htb]
\centering
\includegraphics[width=0.41\textwidth]{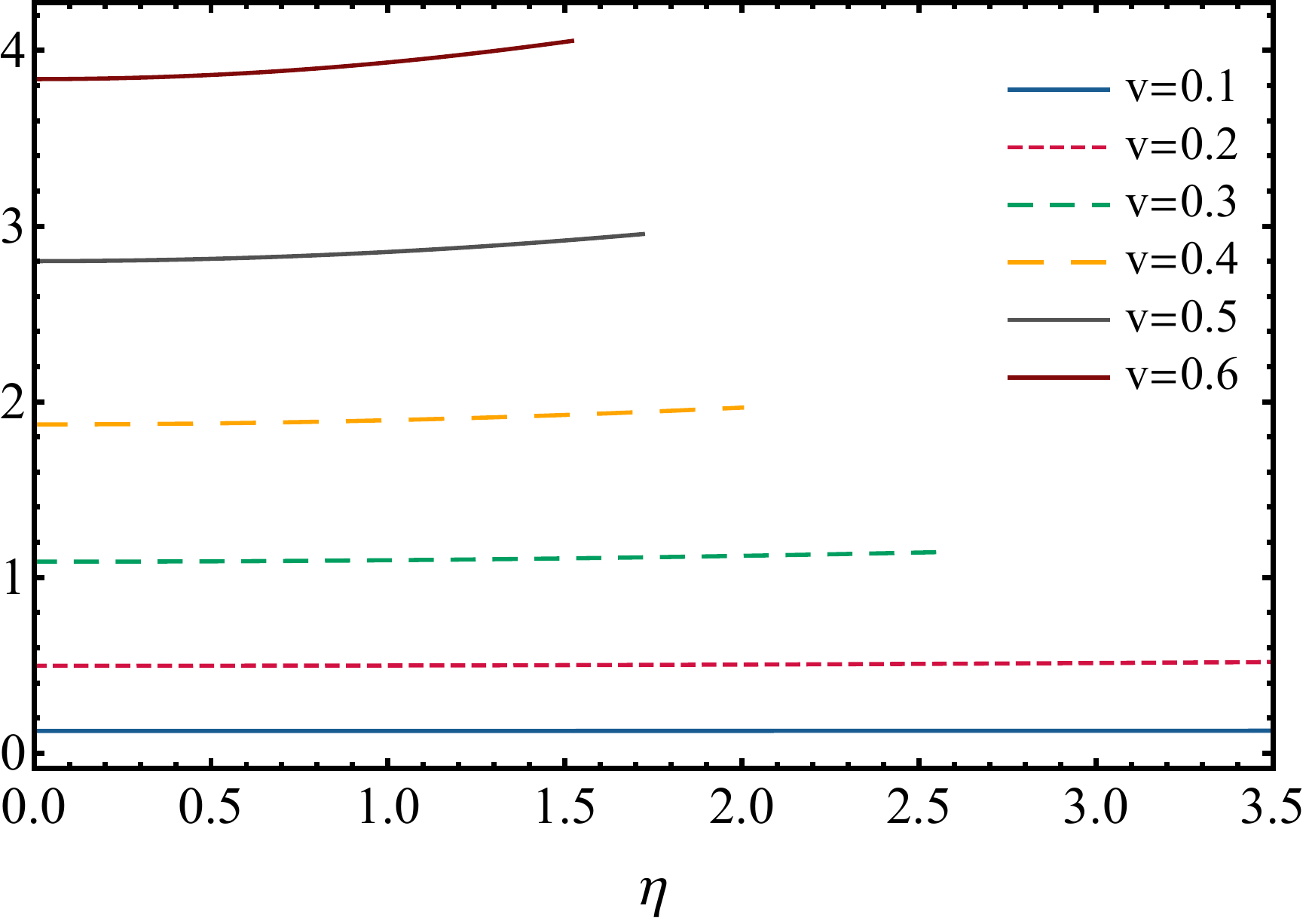} \label{df:1a}\,
\includegraphics[width=0.42\textwidth]{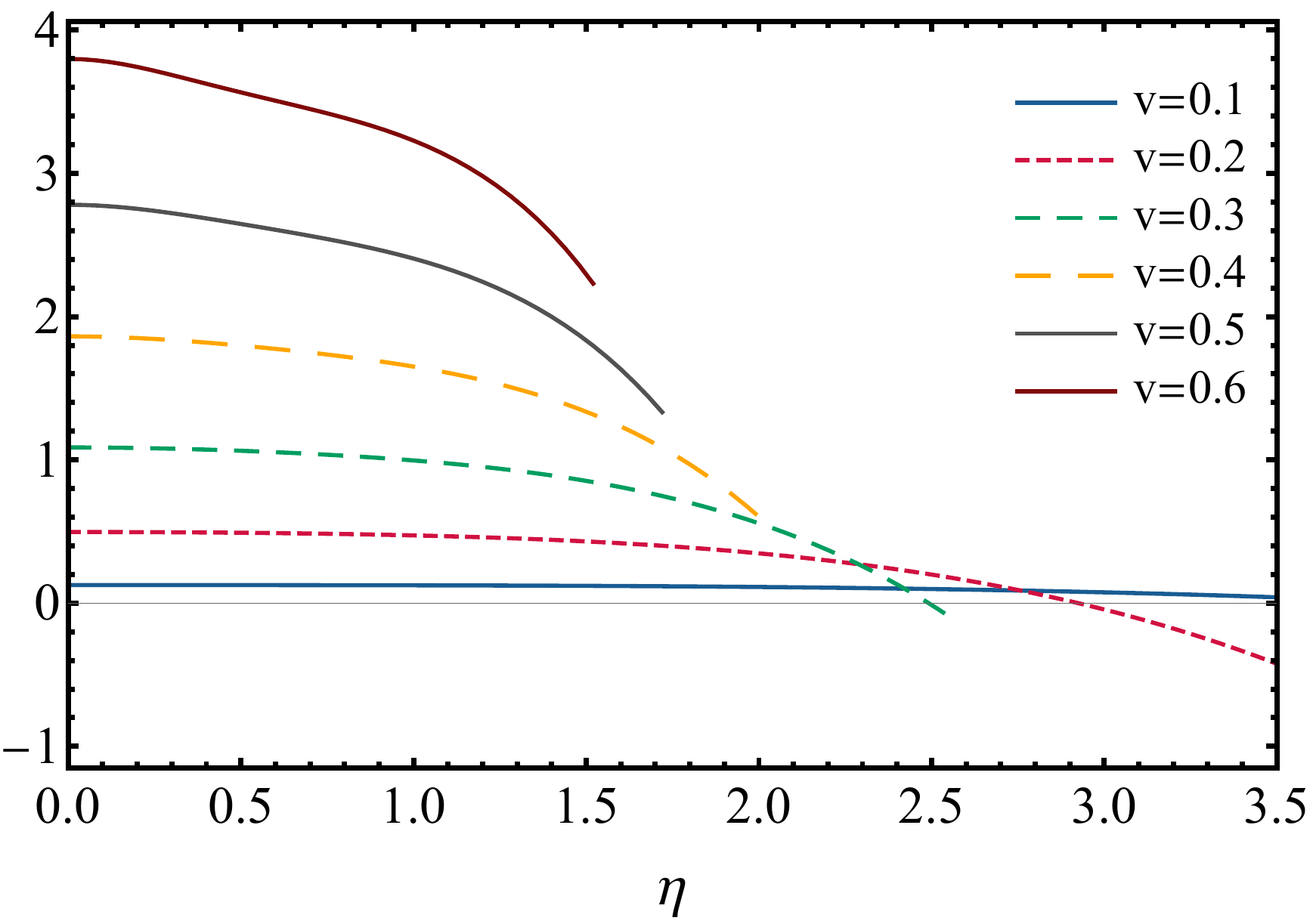} \label{df:1p}
\caption{Percentage deviation of QNM frequencies for the real part of the axial (left) and polar (right) EM QNMs for $l=1$, as derived from the $\mathcal{O}(v^2)$ true equations compared to the QNMs computed using the DF approximation. As expected, the DF approximation works better when $v\to 0$.}
\label{plot:df:1}
\end{figure*}

Using the DF approximation, we computed the $l=1$ EOM at $\mathcal{O}(v^2)$ for the system of coupled scalar and EM fields and computed their QNM spectrum\footnote{The EOM can be found in a supplemental {\scshape Mathematica}\textsuperscript{\textregistered} notebook~\cite{webpages}.}. In Fig.~\ref{plot:df:1} we plot the relative percentage difference for the real part of the $l=1$ EM QNM frequencies between the DF approximation and the exact QNMs (similar results also hold for the imaginary part). The error due to DF approximation is almost negligible for small $v$ and remains quite accurate even for $v\approx 0.6$, i.e. when we already expect the DF approximation to break down. Moreover, the difference is not very sensitive to the particular value of $\eta$. Similar results also hold for the gravitational and EM $l=2$ QNMs. 

\begin{figure}[htb]
\centering
\includegraphics[width=0.42\textwidth]{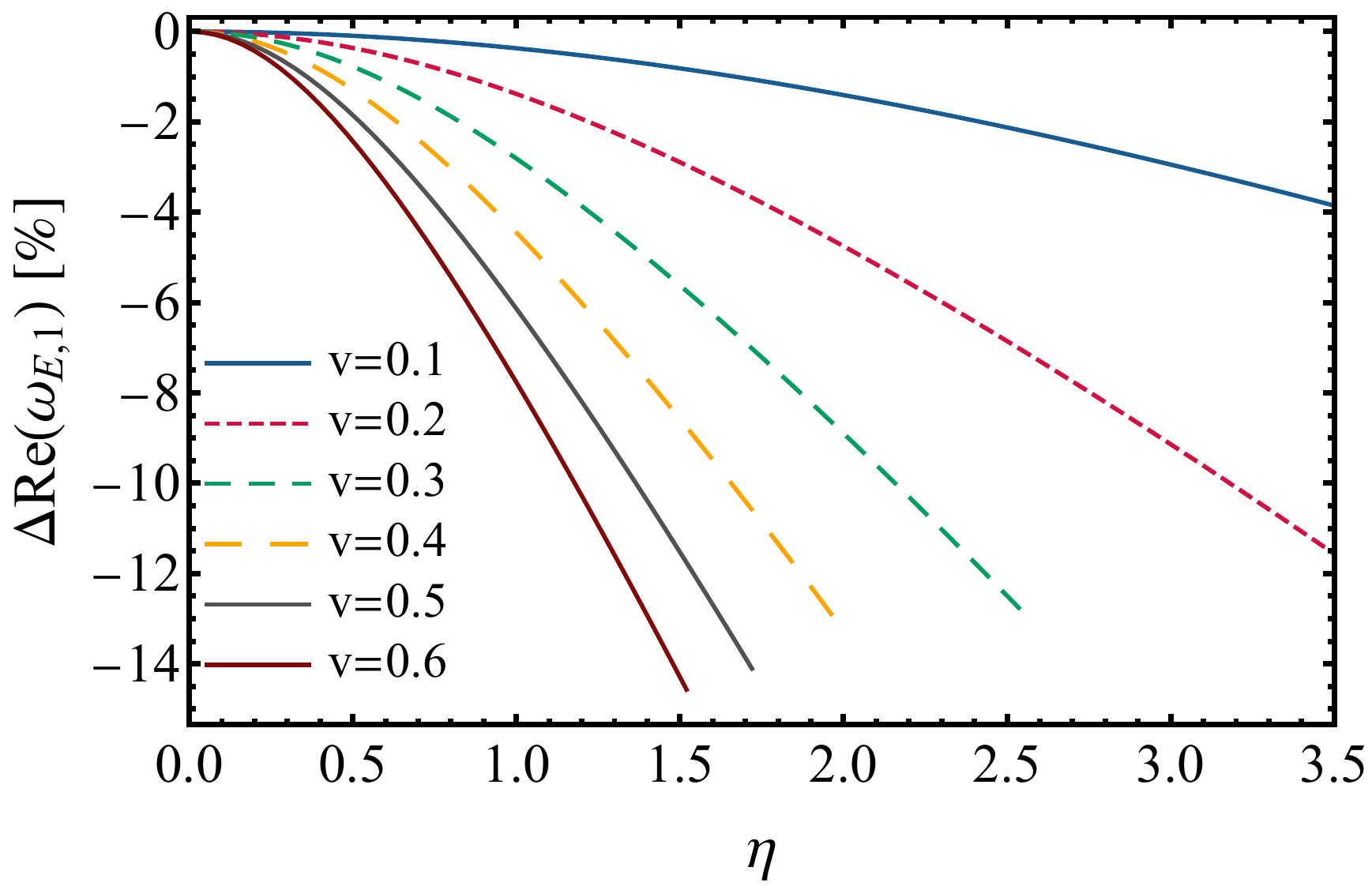}
\caption{ISO-breaking for $l=1$ EM modes as derived from the DF approximate equations (cf. top left panel of Fig.~\ref{plot:delta:e}).}
\label{plot:df:2}
\end{figure}

In Fig.~\ref{plot:df:2} we show the ISO-breaking for the real part of the QNM for $l=1$, as estimated using the DF approximation. By comparison with Fig.~\ref{plot:delta:e}, we can see that the DF prediction remains quite accurate even for $v\approx 0.6$. We thus conclude that the DF approximation captures the main qualitative and quantitative features of the QNM spectrum of EMD BHs, under the approximation of weak charge. In particular, it allows a computationally simpler study of ISO-breaking in the EM channel. In the next subsection we will therefore rely on the DF approximation to compute EM QNMs in the presence of slow rotation.
\subsection{Inclusion of slow rotation}
\label{sec:rotation}
To derive the perturbed EOM for slowly rotating BHs we follow the procedure described in~\cite{pani_review,cardoso_slow_1, berti_slow_1, Brito:2013wya}. In Ref.~\cite{cardoso_slow_1} it was shown that, at linear order in the spin, the radial and angular components of the perturbations are separable, axial and polar modes decouple and the couplings between different multipoles do not affect the QNM frequencies. The resulting equations, which can be found in the supplemental {\scshape Mathematica}\textsuperscript{\textregistered} notebook~\cite{webpages}, are sufficiently similar to the static ones to be addressed with the same techniques. The computation of the QNMs thus proceeds along the same lines of Sec.~\ref{sec:qnm}, the only difference being that the asymptotic behavior of the wave functions reads~\cite{cardoso_slow_1,berti_slow_2}
\be
\label{eq:asy:rot}
Z(r)\sim
\begin{cases}
e^{i\omega\,r_\star} & \text{for $r\to\infty$}\,,\\
e^{-i(\omega-m\Omega_H)\,r_\star} & \text{for $r\to R_+$}\,.
\end{cases}
\ee
Here $m$ is the azimuthal number of the spherical harmonics and $\Omega_H$ is the angular velocity of the BH event horizon
\be
\label{eq:omega:h}
\Omega_H=-\left.\frac{g_{t\phi}}{g_{\phi\phi}}\right|_{r=r_H}=\frac{a\,\Omega(R_+)}{R_+^2\,g(R_+)}=\tilde{a}\left(\frac{2+v^2}{8M}\right)+\mathcal{O}(v^3)\,.
\ee
At first order in the spin we can expand the QNM frequencies $\omega_{l,m}$ in $\tilde{a} m$~\cite{cardoso_slow_1}:
\be
\label{eq:qnm:a}
\omega_{l,m}=\omega_l^{(0)}+\tilde{a}m\,\omega_l^{(1)} + \mathcal{O}(\tilde{a}^2)\,,
\ee
where $\omega^{(0)}_l$ is the frequency of the static BH, while $\omega_l^{(1)}$ is the first order correction to the QNM frequency due to the BH spin. The quantity $\omega_l^{(1)}$ only depends  on the multipole number $l$, the dilaton coupling $\eta$, and on the BH mass and electric charge, while the dependence on $\tilde{a}$ and $m$ factors out at first order. Therefore the computation of the slow-rotation QNMs reduces to the determination of $\omega^{(1)}_l$.

This approximation was used in Ref.~\cite{cardoso_slow_1} to compute the EM QNMs in a slowly rotating Kerr BH background, while Refs.~\cite{berti_slow_1, berti_slow_2} used it to compute the QNMs of Kerr-Newman BHs. In particular, they found that the $\mathcal{O}(a)$ approximation predicts QNM frequencies that deviate from their exact values by less than $1 \%$ for $a\lesssim 0.3$ and $3\%$ when $a\lesssim0.5$. Within this error, they also showed that axial and polar sectors are still isospectral even when including spin. 

Here we extend these computations for the slowly-rotating EMD BHs described by the metric~\eqref{eq:metric:slow}, although limiting our analysis to the weak-charge limit. When $v\lesssim 0.6$ and $\eta=0$, the results of~\cite{berti_slow_1,berti_slow_2} coincide with ours. For concreteness let us focus on the gravitational and the EM modes since the behavior for the scalar QNMs is completely analogous. 

\begin{figure*}[htb]
\centering
\includegraphics[width=0.42\textwidth]{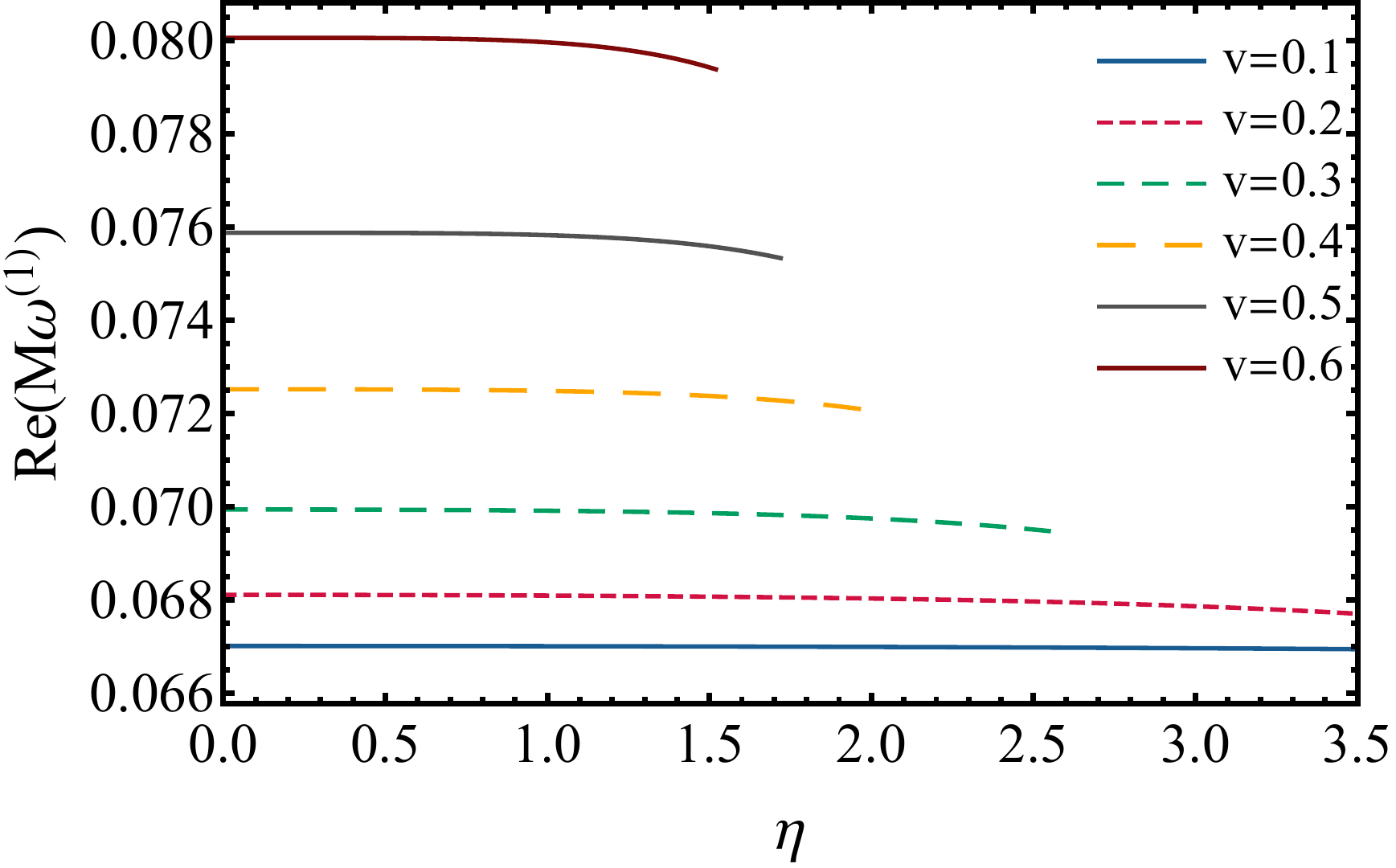}\,
\includegraphics[width=0.44\textwidth]{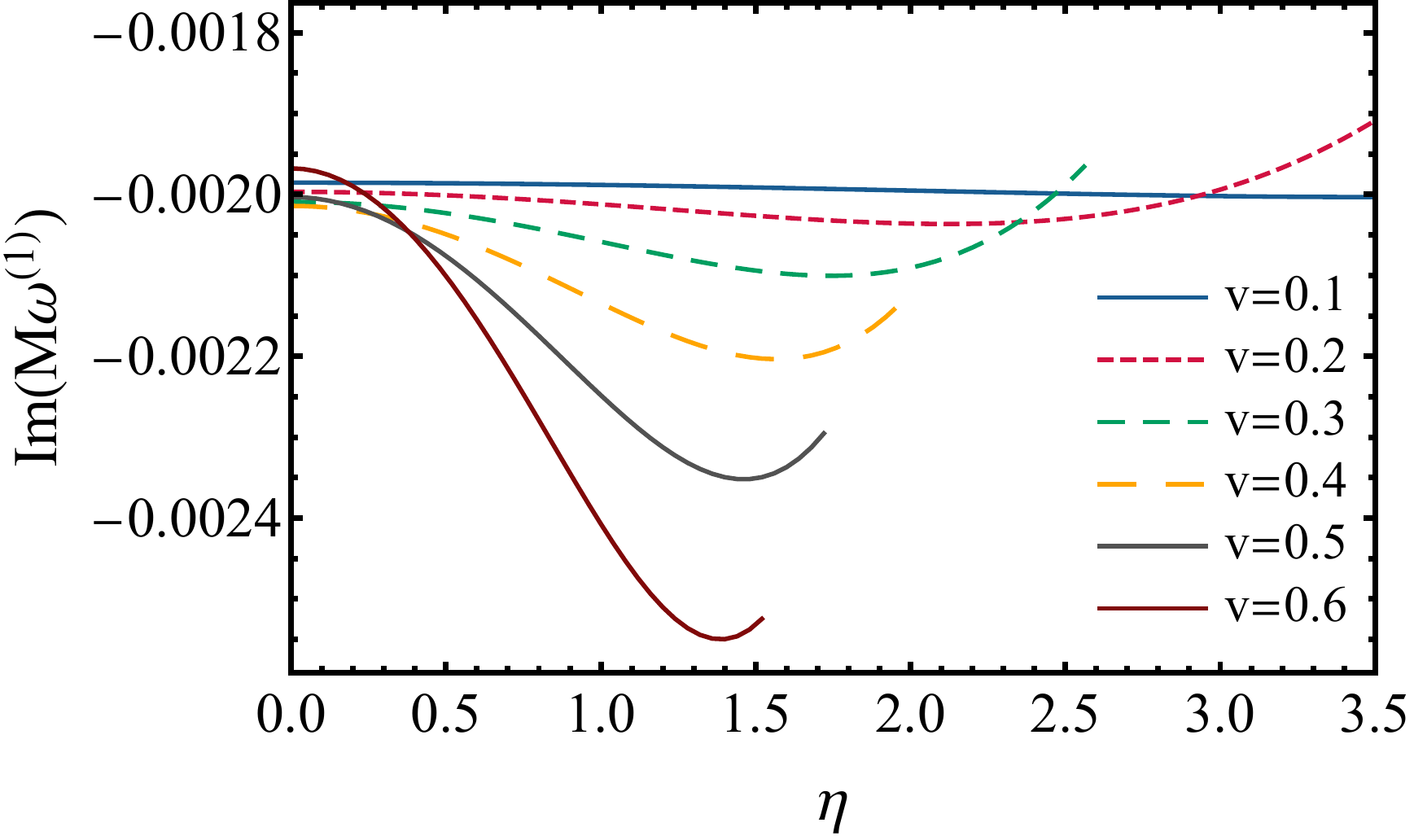}
\caption{Real (left) and imaginary (right) part of the leading-order spin corrections, $M\omega^{(1)}_l$, for the $l=2$ axial gravitational QNMs. Similar results can be obtained for $l>2$.}
\label{plot:ax:slow:2}
\end{figure*}

\paragraph*{Gravitational modes.}
We start by computing the axial gravitational QNMs for $l=2$ (similar results apply to $l>2$). In Fig.~\ref{plot:ax:slow:2} we show the real part of $\omega^{(1)}_2$. When $\eta=0$, these results are in good agreement with the ones plotted in Fig.1 of Ref.~\cite{berti_slow_1} where the QNMs of Kerr-Newman were computed within the slowly-rotating approximation but without any approximation for the BH charge. As in the static case, the dependence on $\eta$ is weak and the modes are very close to those of a Kerr-Newman BH in Einstein-Maxwell.

It is instructive to compare the results of Fig.~\ref{plot:ax:slow:2} with the light ring approximation~\eqref{eq:omega:eikonal:2}. As we already observed, Eq.~\eqref{eq:omega:eikonal:2} is in agreement with the fact that the correction to the imaginary part of the QNM frequency depends very weakly on the spin in the small-charge approximation. Moreover, we also see that the leading-order correction  due to $\tilde{a}$ for the real part of the QNM  ranges from $0.074$ for $v=0.1$ to $0.087$ for $v=0.6$, yielding quite accurate results when compared with Fig.~\ref{plot:ax:slow:2}. Overall we find that Eq.~\eqref{eq:omega:eikonal:2} predicts the $l=2$ gravitational QNM complex frequencies with relative errors always smaller than $\sim 5\%$ for the real part and $\sim 8\%$ for the imaginary part, within the parameter space we consider.

\begin{figure*}[htb]
\centering
\includegraphics[width=0.42\textwidth]{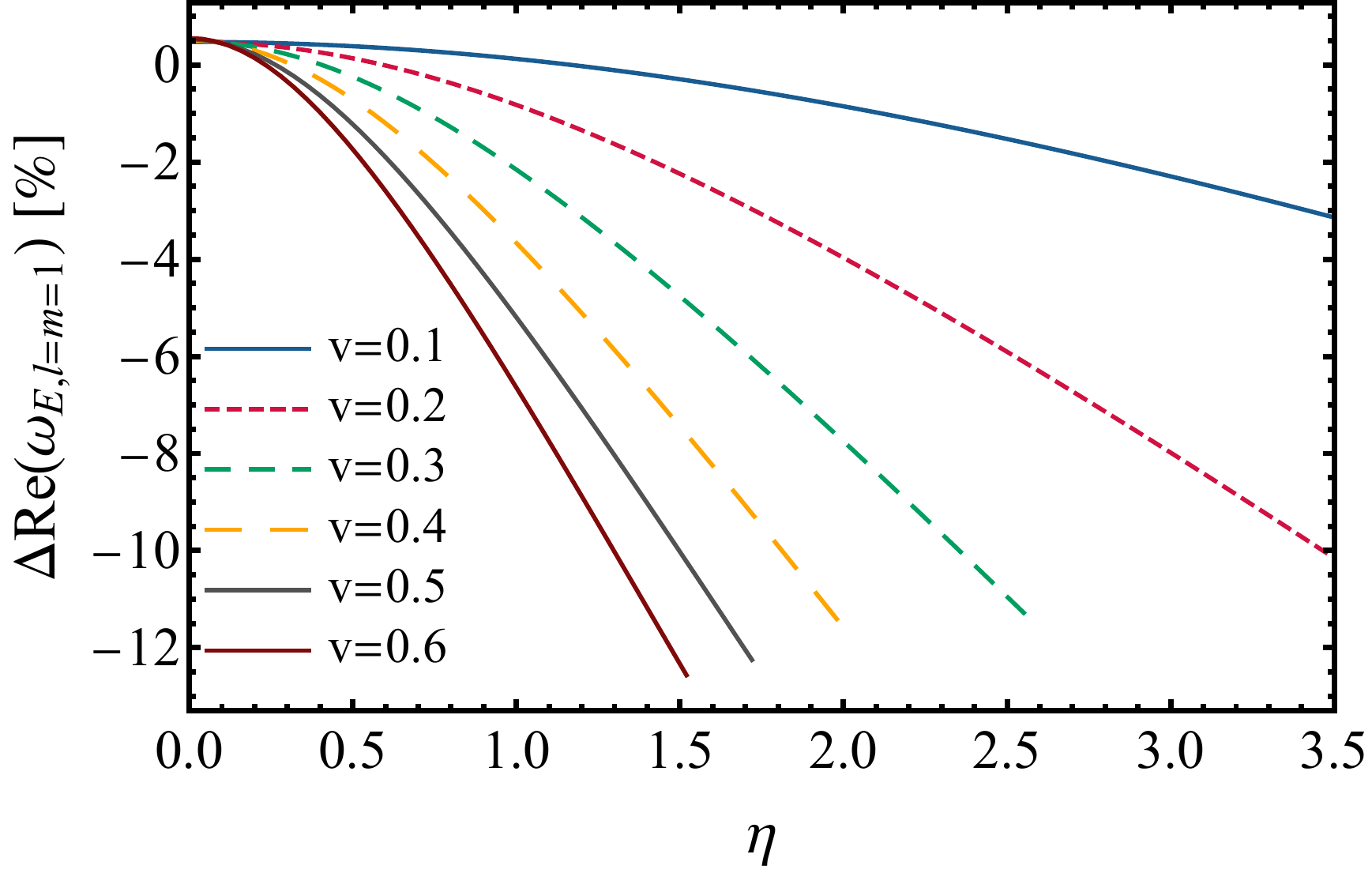}\,
\includegraphics[width=0.42\textwidth]{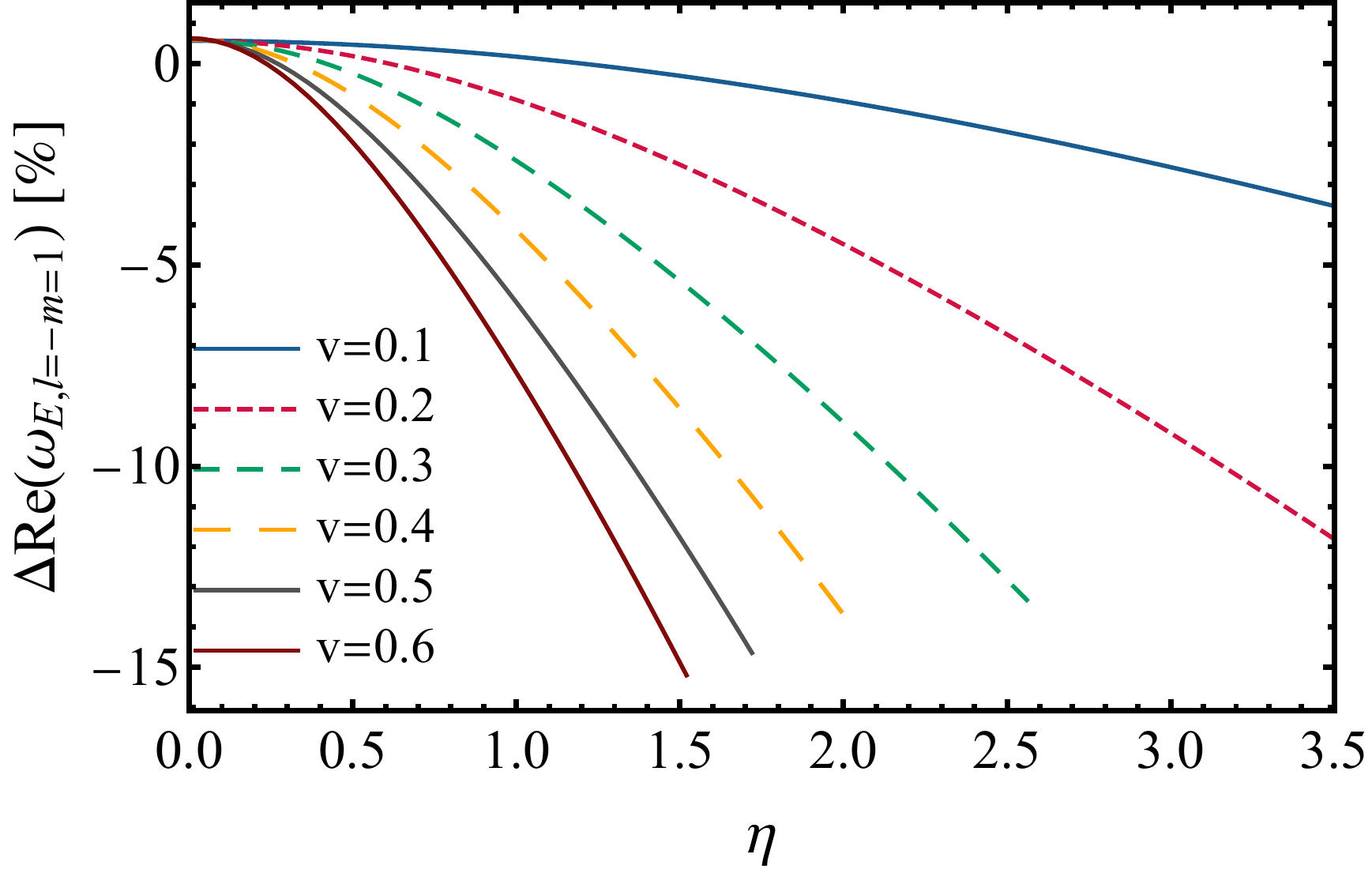}

\caption{ISO-breaking of the real part of the EM QNMs for $l=1$, $m=1$ (left), $m=-1$ (right) and $\tilde{a}=0.2$ [cf. top left panel of Fig.~\ref{plot:delta:e}].}
\label{plot:iso:slow}
\end{figure*}

The polar equations are rather cumbersome to treat. However, guided by the intuition of the static case and the results in Refs.~\cite{berti_slow_1,berti_slow_2}, we expect that the difference with the axial modes will be small.
\paragraph*{Electromagnetic modes.}
It is perhaps more interesting to investigate the difference between axial and polar modes in the EM spectrum, to see how our conclusions in Sec.~\ref{sec:qnm:em} are modified. To this aim, we simplify the problem using the DF scheme, as explained in Sec.~\ref{sec:df}. We concentrate on the real part of the QNMs because it displays the larger effects. Fig.~\ref{plot:iso:slow} shows the EM ISO-breaking for $l=1$, $\tilde{a}=0.2$ and $m=\pm1$ (when $m=0$, Eq.\eqref{eq:qnm:a} implies that the spectrum is unchanged). It is clear from a comparison with Fig.~\ref{plot:delta:e} that the spin does not substantially change the degree of ISO-breaking.
\section{Conclusions}
\label{secc:end}


In this paper we studied the QNMs of weakly charged static and slowly-rotating black holes in EMD. We considered generic values of the dilaton coupling $\eta$, thus extending the analysis of \cite{ferrari_1}, which was restricted to static BHs and the particular case $\eta=1$ and of~\cite{Konoplya:2001ji} who only studied the axial QNMs of static solutions with generic $\eta$.
 
We have shown that, within the parameter space we considered, the gravitational QNMs are very weakly dependent on $\eta$, in agreement with the results of~\cite{Konoplya:2001ji,lehner_2}. On the other hand, the EM QNMs exhibit a clear dependence on $\eta$, mainly visible in the fact that the isospectrality between the polar and the axial sectors can be significantly broken for large enough $\eta$, unlike the gravitational QNMs. By using an approximate treatment of the perturbed equations, inspired by the Dudley-Finley approach, we have shown that this is due to the interaction between the dilaton and the EM field.

We also compared the above results with the light ring approximation. In electrovacuum GR, it is known that the dominant gravitational QNMs can be estimated from the properties of unstable light rings, with a surprisingly good accuracy. While this approach does not capture all the expected properties in generic modified theories of gravity, it should still provide reliable estimates when the deviations from Kerr are parametrized by small perturbative quantities~\cite{post_kerr}. Indeed, we found that, in the weak-charge approximation, the light ring approximation captures the main qualitative features of the gravitational QNMs, also providing a quite accurate quantitative estimate of both the static contribution and the slow-rotation corrections. On the other hand, it does not accurately describe the spectrum of EM and scalar modes, mainly due to the coupling of the EM sector with the dilaton. Our results therefore highlight what is possibly a generic limitation of this approach in theories beyond GR, namely the fact that (i) it fails to predict the existence of new families of QNMs when non-gravitational degrees of freedom are present in the theory, and (ii) it fails to predict the ISO-breaking of axial and polar QNMs. However, at least in the present case, we showed that it is still possible to account for these properties in a simplified way, by means of an hybrid approach in which one (i) estimates the gravitational modes using the light ring correspondence and (ii) treats the matter perturbations in a suitable approximation scheme, neglecting the effect of the metric backreaction.

This work can be improved in several fronts. An obvious extension would be to relax our small charge approximation, to allow for generic BH charges. We expect this to be a rather cumbersome calculation since the perturbation equations become easily intractable for generic BH charge; however some estimates could perhaps be obtained by extending the numerical simulations of Ref.~\cite{lehner_2} to large charges. The other obvious extension would be to relax the small spin approximation. Since exact spinning solutions are only known for the particular case $\eta=\sqrt{3}$~\cite{frolov_metric,solution_3,lehner_1}, obtaining the QNM spectrum for generic $\eta$ and BH spins is most likely only doable with numerical relativity simulations. For the exact spinning solution when $\eta=\sqrt{3}$, we expect that one could obtain the QNM spectrum with no approximations by extending the results obtained in Ref.~\cite{Dias:2015wqa} for the Kerr-Newman case. 

Finally, although we focused on Einstein-Maxwell-dilaton BHs, spinning BH solutions are also known in more generic alternative theories of gravity (see e.g.~\cite{Pani:2011gy}). The methods used in Refs.~\cite{pani_review,cardoso_slow_1, berti_slow_1, Brito:2013wya} and in this paper could be easily extended to compute QNMs for spinning BHs in any theory where exact or slowly-rotating solutions are known. We therefore hope that this paper will stimulate further work in these directions.

\section*{Acknowledgments}

We thank Alessandra Buonanno for useful discussions and for getting us interested in this problem.

\begin{appendices}
\section{Derivation of the perturbed EOM}
\label{appendix:1}
In this Appendix we describe the derivation of the perturbed EOM~\eqref{eq:axial:eom} and~\eqref{eq:polar:eom}. The linearized EOM separate naturally into three groups (see for example \cite{cardoso_slow_1,berti_slow_1,Brito:2013wya}). The first group, or ``scalar group'', includes
\begin{subequations}
\begin{align}
& \delta E_{(I)}=A_l^{(I)}\,Y_l^m=0\,, &I=0,1,2,3\\
& \delta J_{(I)}=A_l^{(I)}\,Y_l^m=0\,, & I=4,5\\
&\delta S=A_l^{(6)}\,Y_l^m=0\,,
\end{align}
\end{subequations}
where $I=0,1,2,3,4,5$ is a shorthand notation for $(tt)$, $(tr)$, $(rr)$, $(+)$, $(t)$ and $(r)$ respectively, and we defined
\be
\delta E_{(+)}=\delta E_{\theta \theta}+\frac{\delta E_{\phi\phi}}{\sin^2\theta}\,.
\ee
The second group, or ``vector group'' is given by
\begin{subequations}
\begin{align}
&\delta E_{(L\theta)}=\alpha_l^{(L)}\,\partial_\theta Y_l^m-\beta_l^{(L)}\frac{\partial_\phi Y_l^m}{\sin\theta}=0\,, & L=0,1\\
&\frac{\delta E_{(L\phi)}}{\sin\theta}=\beta_l^{(L)}\,\partial_\theta Y_l^m+\alpha_l^{(L)}\frac{\partial_\phi Y_l^m}{\sin\theta}=0\,, & L=0,1
\end{align}
\end{subequations}
where $L=0,1$ stand for $(t)$ and $(r)$ respectively, and by
\begin{subequations}
\begin{align}
&\delta J_\theta=\alpha_l^{(2)}\,\partial_\theta Y_l^m-\beta_l^{(2)}\frac{\partial_\phi Y_l^m}{\sin\theta}=0\,,\\
&\frac{\delta J_\phi}{\sin\theta}=\beta_l^{(2)}\,\partial_\theta Y_l^m+\alpha_l^{(2)}\frac{\partial_\phi Y_l^m}{\sin\theta}=0\,.
\end{align}
\end{subequations}
Finally the third group, or ``tensor group'', consists of
\begin{subequations}
\begin{align}
&\frac{\delta E_{\theta\phi}}{\sin\theta}=s_l\frac{X^l}{\sin\theta}+t_l\,W^l=0\,,\\
&\delta E_{(-)}=-t_l\frac{X^l}{\sin\theta}+s_l\,W^l=0\,,
\end{align}
\end{subequations}
where we defined
\be
\delta E_{(-)}=\delta E_{\theta \theta}+\frac{\delta E_{\phi\phi}}{\sin^2\theta},
\ee
and
\begin{subequations}
\label{eq:x:w}
\begin{align}
&X^l=2\partial_\phi(\partial_\theta Y_l^m-\cot\theta Y_l^m)\,,\\
&W^l=\partial^2_\theta Y_l^m-\cot\theta\partial_\theta Y_l^m-\frac{\partial^2_\phi Y_l^m}{\sin^2\theta}\,.
\end{align}
\end{subequations}
The functions $A_l^{(I)}$, $\alpha_l^{(L)}$, $\beta_l^{(L)}$, $s_l$ and $t_l$ are \emph{purely radial} functions of the linearized perturbation fields. The angular dependence of the linearized EOM can be separated using the orthogonality properties of the scalar, vector and tensor spherical harmonics. The result is:
\begin{subequations}
\begin{align}
& A_l^{(I)}=0\,,&\text{$I=0,\dots,6$}\\
&\alpha_l^{(L)}=\beta_l^{(L)}=0\,, &\text{$L=0,1,2$}\\
&s_l=t_l=0\,.
\end{align}
\end{subequations}
Moreover, axial and polar perturbations decouple as
\be
\label{eq:axial:1}
\text{Axial}=
\begin{cases}
&\beta_l^{(L)}=0\quad \text{$L=0,1,2$}\\
& t_l=0
\end{cases}
\ee
\be
\label{eq:polar:1}
\text{Polar}=
\begin{cases}
&A_l^{(I)}=0\quad \text{$I=0,\dots,6$}\\
&\alpha_l^{(L)}=0\quad \text{$L=0,1,2$}\\
&s_l=0
\end{cases}
\ee
We consider the two cases separately.\\
\paragraph{Axial perturbed equations.}
\label{sec:small:axial}
The axial EOM \eqref{eq:axial:1} can be reduced to two second order differential equations. We briefly describe the procedure. From \eqref{eq:metric:pert} and \eqref{eq:vector:pert}, the axial perturbation fields are $h_0(r)$, $h_1(r)$ and $u_4(r)$. It is convenient to make the redefinitions
\be
\label{eq:redef:axial}
h_1(r)=\frac{Q_1(r)}{f(r)}\,,\qquad u_4(r)=\frac{i}{\omega}U_4(r)\,.
\ee
From $t_l=0$ we get $h_0(r)=if\,Q_1'(r)/\omega$, thus eliminating $h_0(r)$ from the system \eqref{eq:axial:1}. $\beta_l^{(0)}=0$ is implied by $\beta_l^{(1)}=0$ and $\beta_l^{(2)}=0$, which thus are the only two remaining independent equations. The system is further simplified by replacing $\beta_l^{(2)}=0$ with the linear combination
\be
\label{eq:redef:axial:2}
\beta_l^{(2)}+\left(\frac{2Mv\,f(r)}{r^2\omega}\right)\beta_l^{(1)}=0\,.
\ee
Next, solving the system for $Q_1''(r)$ and $U_4''(r)$ and making the further change of variables
\be
\label{eq:redef:axial:3}
Q_1(r)=r\sqrt{g(r)}\,\hat{Q}(r)\,,\qquad U_4(r)=\sqrt{g(r)}\,\hat{U}(r)\,\,
\ee
we obtain the system
\be
\label{eq:axial:final}
\left[\frac{d^2}{dr\indices{_\star^2}}+\omega^2\right]
\begin{pmatrix}
\hat{Q}(r)\\
\hat{U}(r)
\end{pmatrix}
=\mathbb{V}^A
\begin{pmatrix}
\hat{Q}(r)\\
\hat{U}(r)
\end{pmatrix}\,,
\ee
where $\mathbb{V}^A$ is a nondiagonal $2\times 2$ matrix. The system \eqref{eq:axial:final} can be diagonalized by an $r$-independent linear transformation, thus reducing to \eqref{eq:axial:eom:2}.\\\
\paragraph{Polar perturbed equations.}
The treatment of the polar perturbation equations \eqref{eq:polar:1} is more complicated. The polar perturbation fields are $H_0(r)$, $H_1(r)$, $H_2(r)$, $K(r)$, $f_{01}(r)$, $f_{01}(r)$, $f_{12}(r)$ and $z(r)$. It is convenient to make the redefinitions
\be
\label{eq:redef:polar}
f_{12}(r)=\frac{i\omega\,g(r) F_1(r)}{f(r)}\,,\qquad H_1(r)=\omega R_1(r)\,.
\ee
The equations $s_l=0$, $\alpha_l^{(2)}=0$ and $A_l^{(5)}=0$ can be used to eliminate $H_0(r)$, $f_{02}(r)$ and $f_{01}(r)$ respectively. Using the Bianchi identity \eqref{eq:bianchi} we obtain the Maxwell equation, a second order differential equation for $F_1(r)$. $A_l^{(6)}=0$ is the scalar equation, a second order differential equation for $z(r)$.

Among the remaining equations, the only independent ones are $A_l^{(1)}=0$, $A_l^{(2)}=0$, $\alpha_l^{(0)}=0$ and $\alpha^{(1)}_l=0$. They are first order differential equations in the gravitational and matter perturbations. Following Zerilli \cite{zerilli_1,zerilli_2}, we solve the system $\{A_l^{(1)}=0,\alpha_l^{(0)}=0,\alpha_l^{(1)}=0\}$ for $\{K'(r),R_1'(r),H_2'(r)\}$, we plug the solution into $A_l^{(2)}=0$ and we solve for $H_2(r)$, thus eliminating $H_2(r)$ from the system. 

We are left with two coupled first order differential equations for $K(r)$ and $R_1(r)$. As shown by Zerilli \cite{zerilli_1,zerilli_2}, they can be reduced to a single second order differential equation. To this purpose we introduce two new gravitational variables $\hat{K}(r)$ and $\hat{R}(r)$, related to the original ones by a linear transformation:
\begin{subequations}
\label{eq:zerilli:redef}
\begin{align}
&K(r)=\alpha(r)\hat{K}(r)+\beta(r)\hat{R}(r)\,,\\
&R_1(r)=\gamma(r)\hat{K}(r)+\lambda(r)\hat{R}(r)\,.
\end{align}
\end{subequations}
The procedure consists of choosing the functions $\alpha(r),\beta(r),\gamma(r)$ and $\lambda(r)$ such that the gravitational perturbation equations assume the form
\begin{subequations}
\begin{align}
&\frac{d\hat{K}(r)}{dr_\star}=\hat{R}(r)+\text{(matter couplings)}\label{eq:k}\,,\\
&\frac{d\hat{R}(r)}{dr_\star}+\omega^2\hat{K}(r)=V_K(r)\hat{K}(r)+\text{(matter couplings)}\label{eq:r}\,,
\end{align}
\end{subequations}
where $V_K(r)$ is a potential, and the matter couplings refer to terms linear in $F_1(r)$, $z(r)$ and their first derivatives. Eqs. \eqref{eq:k} and \eqref{eq:r} can then be combined into a single second order equation,
\be
\label{eq:k:2}
\left[\frac{d^2}{dr\indices{_\star^2}}+\omega^2\right]\hat{K}(r)=V_K(r)\hat{K}(r)+\text{(matter couplings)}\,.
\ee
In practice, the procedure translates into three algebraic and one differential equations. The three algebraic equations allow us to express $\alpha(r)$, $\gamma(r)$ and $\lambda(r)$ as linear functions of $\beta(r)$. The differential equation is a first order differential equation for $\beta(r)$.

The algebraic equations must be treated carefully. Indeed, since we are working at $\mathcal{O}(v^2)$, we cannot restrict to exact solutions, but we must also allow for solutions valid at $\mathcal{O}(v^2)$, thus enlarging the space of the admitted solutions. In fact, we find that we must use appropriate linear combinations of the exact solutions. We derived the correct coefficients working in the special cases $\eta=0,1,\sqrt{3}$, for which we were able to derive the expressions for $\alpha(r)$,$\gamma(r)$ and $\lambda(r)$ at all orders in $v$. Since the coefficients are independent of $\eta$, we assumed their validity for generic values of $\eta$ and verified \emph{a posteriori} that we obtain a consistent $\mathcal{O}(v^2)$ solution for $\alpha(r)$, $\beta(r)$, $\gamma(r)$ and $\lambda(r)$ for all $\eta$. In particular
\be
\label{eq:beta}
\beta(r)=1-\frac{\eta^2Mv^2}{r}\equiv g(r)\,.
\ee
We can now write the Maxwell, scalar and gravitational EOM. The Maxwell and scalar EOM are more conveniently expressed in the new variables $\hat{F}(r)$ and $\hat{S}(r)$ defined by
\begin{subequations}
\label{eq:zerilli:redef:2}
\begin{align}
&F_1(r)=\hat{F}(r)/\sqrt{g(r)}\,,\\
&z(r)=\hat{S}(r)+\frac{2\,\eta\,Mv}{r}\hat{F}(r)\,.
\end{align}
\end{subequations}
The final result is the system of coupled equations \eqref{eq:polar:eom}.
\section{Approximation error at $\mathcal{O}(v^2)$}
\label{appendix:2}
\begin{figure*}[htb]
\centering
\includegraphics[width=0.42\textwidth]{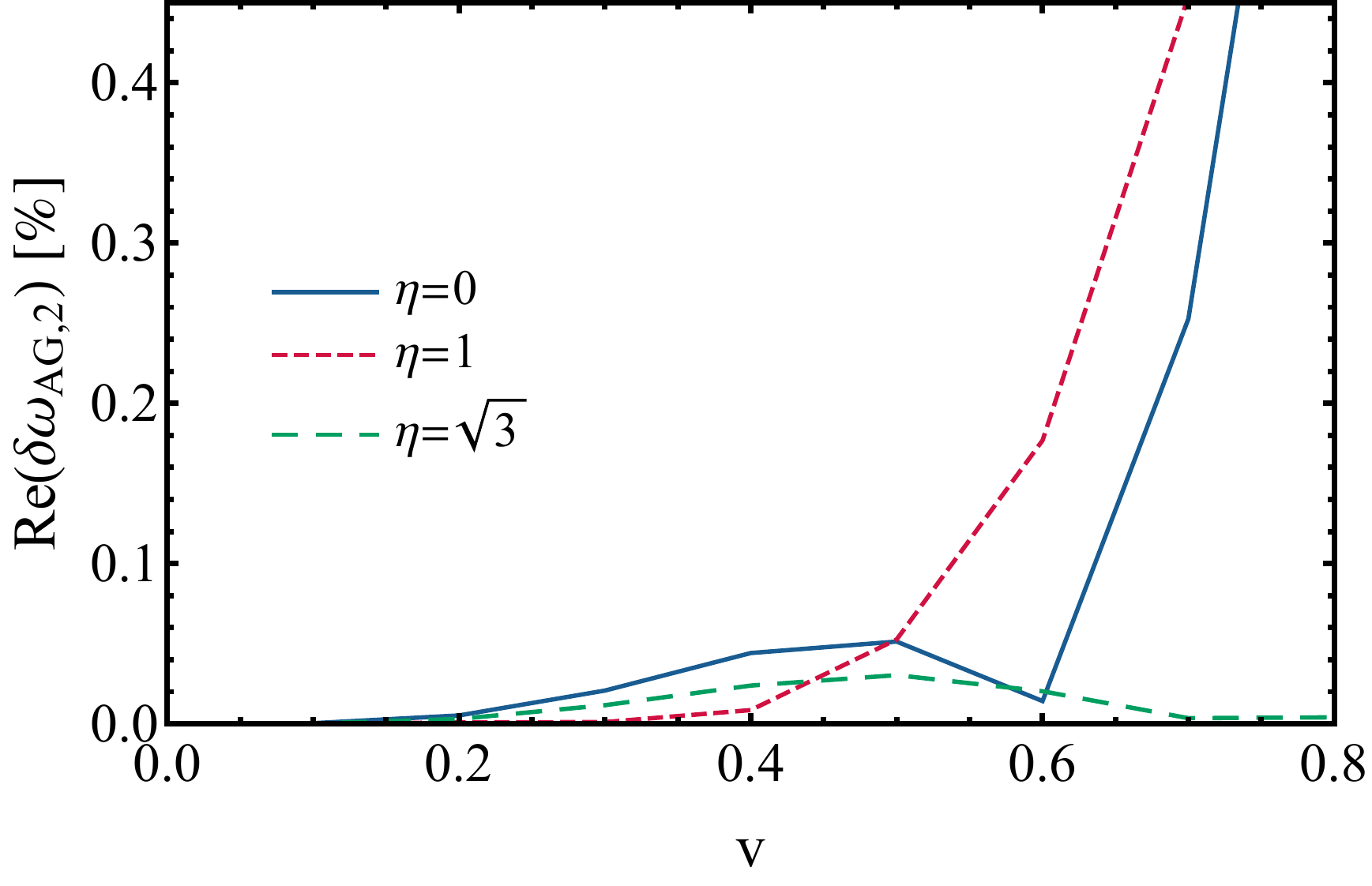}\,
\includegraphics[width=0.42\textwidth]{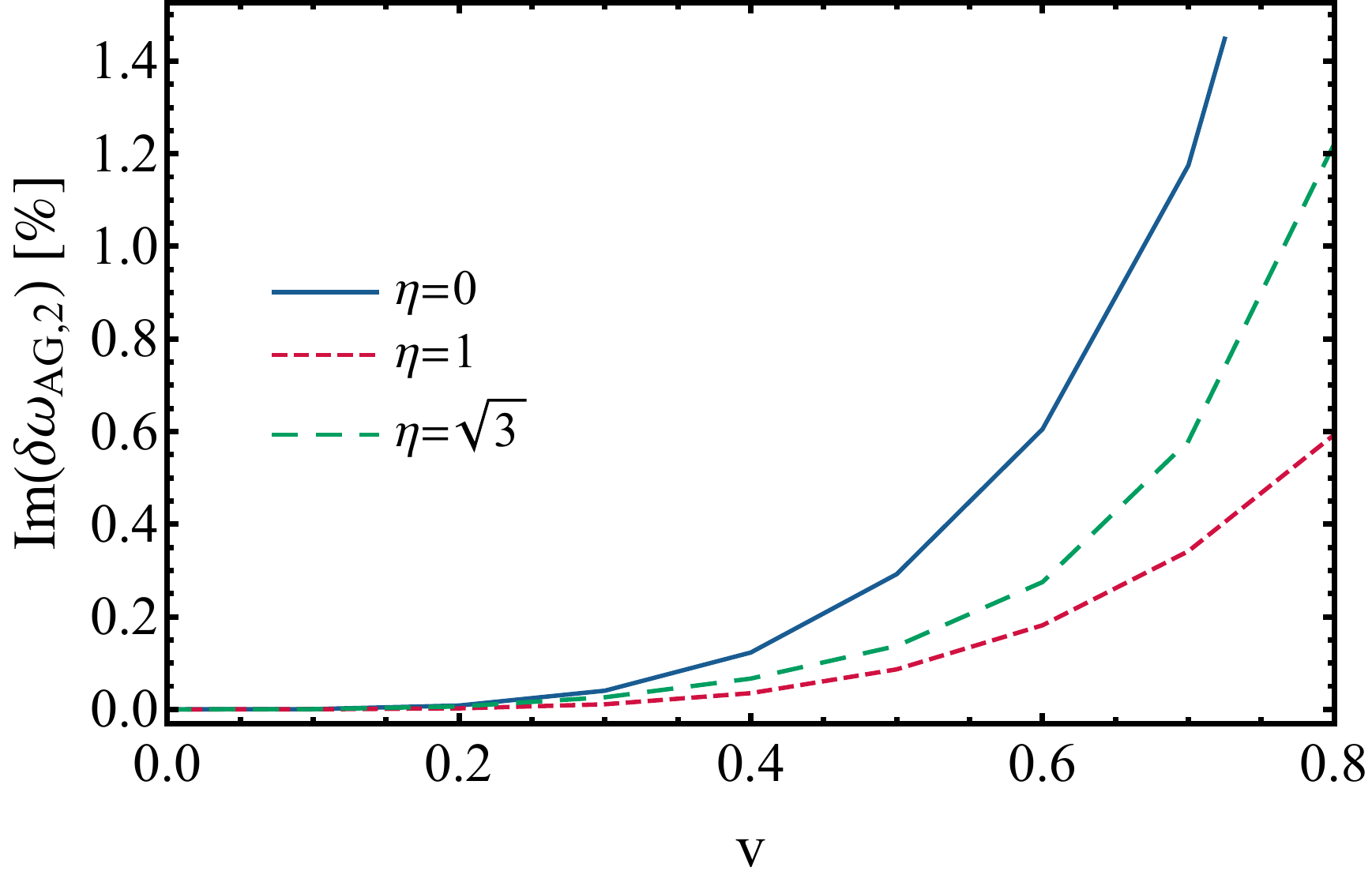}
\caption{Real (left) and imaginary (right) part of $\delta\,\omega_{AG}$ for $l=2$, as defined in Eq. \eqref{eq:delta:full}, as a function of the BH's charge-to-mass-ratio, $v$, for different values of $\eta$.}
\label{plot:delta:full}
\end{figure*}
Following the same procedure as in Appendix \ref{appendix:1}, we derived the axial perturbed EOM at all orders in $v$ in a diagonalized form for the static BH. They have the same form of Eqs.\eqref{eq:axial:eom}, but the potentials are now given by
\be
\label{eq:full:v}
V_{1,2}^A(r)=\frac{(r-R_+) \left(1-\frac{R_-}{r}\right)^{-4 \eta ^2/(\eta ^2+1)}}{2 \left(\eta ^2+1\right)^2 r^6}\left(X_\eta(r)\mp Y_\eta(r)\right)\,,
\ee
where the functions $X_\eta(r)$ and $Y_\eta(r)$ are
\begin{widetext}
\be
Y_\eta(r)=\left(\eta ^2+1\right) r \,(r-R_-) \sqrt{2 \left(\eta ^2+1\right) R_+ R_- \left(-3 \eta ^2+\,8 l(l+1)-7\right)+9 \left(\eta ^2+1\right)^2 R_+^2+\left(\eta ^2-3\right)^2 R_-^2}\,
\ee
and
\begin{multline}
X_\eta(r)=2 \left(\eta ^2+1\right)^2 l(l+1) r^2 (r-R_-) -\left(\eta ^2+1\right) r^2 \left[3 \left(\eta ^2+1\right) R_+ +\left(3-5 \eta ^2\right) R_-\right]+\\
+r\, R_- \left[\left(-3 \eta ^4+8 \eta ^2+11\right) R_+ +\left(\eta ^4-2 \eta ^2+3\right) R_-\right]-2 \left(\eta ^2+4\right) R_+ R_-^2\,.
\end{multline}
\end{widetext}
Notice that the exact axial potentials depend on $v$ only through its square $v^2$. The superscripts $1$ and $2$ refer to the Regge-Wheeler and to the Maxwell equations respectively.

As explained at the end of Sec.\ref{sec:qnm:gr}, we use the exact Regge-Wheeler equation to estimate the error due to the $\mathcal{O}(v^2)$ approximation. For concreteness we focus on the values $\eta=0,1$ and $\sqrt{3}$, and we restrict to $v\leq0.8$. In Fig.\ref{plot:delta:full} we plot the relative percentage difference
\be
\label{eq:delta:full}
\delta\,\text{Re}(\omega)=100\times\left| \frac{\text{Re}(\omega_\text{FULL})-\text{Re}(\omega_\text{SMALL})}{\text{Re}(\omega_\text{FULL})}\right|
\ee
and similarly for the imaginary part, where $\omega_\text{FULL}$ corresponds to the QNM frequency computed without any approximation and $\omega_\text{SMALL}$ the QNM frequency computed at $\mathcal{O}(v^2)$. We see that for the real part the error remains below $0.2\%$ for $v\leq 0.6$, while it becomes of the order of half a percent for the imaginary part. It is reasonable to expect that similar errors will occur also in the EM and scalar sectors.
\section{Instability of the dilaton in the large coupling limit.}
\label{appendix:3}
The potential $V^0(r)$ in \eqref{eq:eom:0}, without restricting to weak charges, has the rather lengthy expression
\begin{widetext}
\begin{multline}
V^0(r)=\left(1-\frac{R_+}{r}\right) \left(1-\frac{R_-}{r}\right)^{\frac{-4\eta ^2}{\eta ^2+1}}\frac{1}{\left(\eta ^2+1\right)^2 r^5\left[r(1+\eta ^2)-R_-\right]^2}\left\{\left(\eta ^2+1\right)^3 r^4 \left[(1+\eta ^2)R_+-(\eta ^2-1)R_-\right]\right.\\
\left.+\left(\eta ^2+1\right)^2 r^3 R_-\left[\left(2 \eta ^2-5\right)\left(\eta^2+1\right) R_+-3 R_-\right]+\left(\eta ^2+1\right) r^2 R_-^2 \left[\left(2 \eta ^4+\eta ^2+3\right) R_--\left(\eta^2+1\right)\left(2\eta^4+\eta^2-9\right) R_+\right]\right.\\
\left.-r R_-^3 \left[\left(2\eta^2+7\right)\left(\eta^2+1\right) R_++R_-\right]+\left(\eta ^2+2\right) R_+R_-^4\right\}\,,
\end{multline}
\end{widetext}
where $R_\pm$ are given in \eqref{eq:rpm:1}. In the limit $\eta\to\infty$, it reduces to
\begin{multline}
\label{v0:infty}
V^0(r)\to\left(1-\frac{R_+}{r}\right)\left(1-\frac{R_-}{r}\right)^{-4}\times\\
\times\frac{\left[r^2(R_+-R_-)+2R_+R_-(r-R_-)\right]}{r^5}\,,
\end{multline}
from which it is clear that $V^0(r)>0$ everywhere for $r>R_+$. One may note that, formally, $R_\pm$ diverge in the limit $\eta\to\infty$. However, this can be fixed by rescaling $v\to\sigma/\eta$. With this rescaling one has
\be
\label{eq:rpm:limit}
R_\pm=M\left(\sqrt{1+\sigma^2}\pm 1\right)\,,
\ee
and \eqref{v0:infty} is positive for any value of $\sigma$.

The same computation can be repeated by only perturbing the dilaton field while keeping the vector field fixed, similarly to what was done in Ref.~\cite{lehner_2}. In this case the potential $V^0_\text{back}(r)$, whose general expression is presented in a supplemental {\scshape Mathematica}\textsuperscript{\textregistered} notebook~\cite{webpages}, in the limit $\eta\to\infty$ reduces to
\begin{multline}
\label{v0:infty}
V^0_\text{back}(r)\to\left(1-\frac{R_+}{r}\right)\left(1-\frac{R_-}{r}\right)^{-4}\times\\
\times\frac{\left[r^2(R_+-R_-)-2R_+R_-(r-R_-)\right]}{r^5}\,,
\end{multline}
and it can be easily checked that this expression is not always positive for $r>R_+$. Indeed, using \eqref{eq:rpm:limit}, one can easily see that the factor $r^2(R_+-R_-)-2R_+R_-(r-R_-)$ can be negative for $\sigma>2\sqrt{2}$. Moreover, a numerical inspection reveals that the integral
\be
\label{v:int}
\mathcal{I}=\int_{R_+}^{\infty}\mathcal{V}(r)dr\,,
\ee
becomes negative for $\sigma\gtrsim3.08$, where $\mathcal{V}=V^0_\text{back}/F(r)$. The negativity of $\mathcal{I}$ is a sufficient criterion for the occurrence of instabilities \cite{bh_scalarization_1,bound_states}. Therefore we conclude that this approximation wrongly predicts unstable EMD BHs. 
\end{appendices}
\end{document}